\documentclass[usenatbib]{mnras}

\usepackage[T1]{fontenc}
\usepackage{ae,aecompl}

\usepackage{graphicx}	
\usepackage{amsmath}	
\usepackage{amssymb}	
\usepackage{mathrsfs}
\usepackage{enumitem}
\usepackage[normalem]{ulem} 
\usepackage{multirow} 
\usepackage{xcolor} 
\usepackage{lipsum} 
\usepackage{hyperref} 
\usepackage{mathtools} 
\usepackage{subcaption}
\usepackage{array} 
\usepackage{lipsum}
\usepackage{txfonts}
\usepackage{float}
\usepackage{physics}
\usepackage[shortcuts]{extdash}
\usepackage{lipsum}  
\usepackage{dblfloatfix}
\usepackage{siunitx}
\usepackage{booktabs}
\usepackage{amstext} 
\usepackage{array}   
\usepackage{tablefootnote}
\newcolumntype{L}[1]{>{\centering\arraybackslash$}p{#1}<{$}}

\newcommand{\Lya}{Ly${\alpha}$}

\newcommand{\TNG}{\mbox{\scshape{IllustrisTNG}}\normalfont}

\DeclareRobustCommand{\VAN}[3]{#2}
\let\VANthebibliography\thebibliography
\def\thebibliography{\DeclareRobustCommand{\VAN}[3]{##3}\VANthebibliography}

\makeatletter
\def\footnoterule{\kern-3\p@
  \hrule \@width 2in \kern 2.6\p@} 
\makeatother

\title[Modeling HI Distributions in FDM Cosmologies]{Modeling Post-Reionization HI Distributions in Fuzzy Dark Matter Cosmologies Using Conditional Normalizing Flows}

\author[T. Dome et al.]
{\parbox[t]{\textwidth}
{Tibor Dome$^{1,2}$\thanks{E-mail: td448@cam.ac.uk},
Rumail Azhar$^{1}$,
Anastasia Fialkov$^{1,2}$
}
\\ \\
$^{1}$Institute of Astronomy, University of Cambridge, Madingley Road, Cambridge, CB3 0HA, UK\\
$^{2}$Kavli Institute for Cosmology, Madingley Road, Cambridge, CB3 0HA, UK
}

\date{Accepted XXX. Received YYY; in original form ZZZ}

\pubyear{2023}

\begin{document}
\label{firstpage}
\pagerange{\pageref{firstpage}--\pageref{lastpage}}
\maketitle

\begin{abstract}
Upcoming 21 cm intensity mapping experiments like the Square Kilometer Array (SKA) hold significant potential to constrain the properties of dark matter. In this work, we model neutral hydrogen (HI) distributions using high-resolution hydrodynamical $N$-body simulations of both cold dark matter (CDM) and fuzzy dark matter (FDM) cosmologies in the post-reionization redshift range of $z=3.42-4.94$. We show that the HI abundance decreases in FDM-like cosmologies. Extreme FDM models with $m\sim 10^{-22}$ eV are at odds with a range of measurements. Due to the increased halo bias, the HI bias increases, paralleled by the damped Lyman-$\alpha$ (DLA) bias which we infer from the cross-section of DLAs. The distribution of the latter in extreme FDM models has a high median at the low-mass end, which can be traced to the high column density of cosmic filaments. FDM models exhibit a very similar abundance of DLAs compared to CDM while sub-DLAs are already less abundant. We study the prospects of detecting the brightest HI peaks with SKA1-Low at $z=4.94$, indicating moderate signal-to-noise ratios (SNR) at angular resolution $\theta_A = 2^{\prime}$ with a rapidly declining SNR for lower values of $\theta_{A}$. After training the conditional normalizing flow network HIGlow on 2D HI maps, we interpolate its latent space of axion masses to predict the peak flux for a new, synthetic FDM cosmology, finding good agreement with expectations. This work thus underscores the potential of normalizing flows in capturing complex, non-linear structures within HI maps, offering a versatile tool for conditional sample generation and prediction tasks.
\end{abstract}

\begin{keywords} methods: numerical - galaxies: high-redshift - galaxies: intergalactic medium - cosmology: dark matter - cosmology: large-scale structure of Universe - radio continuum: general
\end{keywords}



\section{Introduction}
\label{s_intro}
21 cm cosmology studies the redshifted 21 cm-wavelength photons that are emitted in a hyperfine transition of atomic hydrogen atoms (HI). The transition occurs when an electron in the excited spin triplet state flips its spin relative to the proton and falls into the singlet state. While this transition is quantum mechanically `forbidden' to first order, a very large number of neutral hydrogen atoms in massive clouds render the signal observable.\par 

The hyperfine transition can be used to trace the matter field through a technique called 21 cm intensity mapping (IM) \citep{Chang_2008, Wyithe_2008, Santos_2015, Navarro_2015}. It is possible since the spin transition in HI is optically thin \citep{Furlanetto_2006}, meaning that 21 cm photons are unlikely to be absorbed/emitted more than once as they travel to our telescopes. Therefore, full 3D mappings of the 21 cm line become possible, with the redshift of the signal providing information about the line of sight distance. Unlike galaxy redshift surveys, there is no need to resolve individual galaxies, which is often required only to determine their redshift. By mapping the unresolved emission of all HI at each frequency of observation and using the observed frequency-redshift relation $\nu= \nu_{21}/(1+z)$, we can directly map to the corresponding redshift. This is further facilitated by the ease at which spectral resolution can be obtained in radio astronomy.\par

Ongoing and upcoming 21 cm IM experiments such as HIRAX
\citep{Newburgh_2016}, PUMA \citep{Bandura_2019}, CHIME \citep[]{CHIME_2022} and \href{https://www.skao.int/}{SKA} \citep{Combes_2021} are set to yield a wealth of insights into cosmology and astrophysics. With minimal angular resolution requirements and expansive fields of view, these instruments can efficiently map vast cosmological volumes. The 21 cm signal in the post-reionization era (and also during reionization) holds great promise for studying alternative dark matter (DM) scenarios. The 21 cm signal is not only sensitive to DM decays, annihilation processes \citep{Liu_2018, List_2020} or interactions between DM and standard model particles \citep{Barkana_2018, Munoz_2018}. It can also probe DM models that result in a suppression of the small-scale matter power spectrum \citep{Sitwell_2014, Schneider_2018, Lidz_2018, Nebrin_2019, Boyarsky_2019, Jones_2021, Bauer_2021}, which are the focus of this paper.\par 

DM models with a small-scale suppression are of interest since the standard model of cosmology and DM - the $\Lambda$CDM model - faces some challenges including the lack of direct detection of DM particles and small-scale observations that are at odds with $\Lambda$CDM predictions \citep{Weinberg_2015, Bullock_2017}. While attempts at unified baryonic solutions to the $\Lambda$CDM small-scale problems exist \citep[as reviewed in][]{Popolo_2017}, here we investigate fuzzy dark matter \citep[FDM]{Hu_2000}, which is made up of ultralight bosonic particles of mass $\sim 10^{-22}$ eV called axions, corresponding to de Broglie wavelengths of $\lambda_{\text{dB}} \sim 1$ kpc. The motivations behind considering multiple species of light axions is vast, encompassing well-established predictions of string/M-theory \citep{Arvanitaki_2010, Demirtas_2018} as well as various field theory extensions of the standard model \citep{Peccei_1977, Kim_2016}. One of the distinctive attributes of FDM is its largely redshift-insensitive comoving de Broglie wavelength $\lambda_{\text{db,c}}$ \citep{Khlopov_1985} which scales as $\lambda_{\text{db,c}} \sim (1 + z)^{1/4}m^{-1/2}$. This unique property simultaneously addresses challenges associated with small-scale structure and constrains the central density of collapsed halos \citep{Schive_2014}, offering a natural resolution to some of the limitations faced by $\Lambda$CDM. Notably, FDM introduces solitonic cores within halo centers, which, in models incorporating axion self-interactions, can undergo a phase transition from dilute to denser states \citep{Mocz_2023}.\par 

In our investigation, we run $N$-body hydrodynamical simulations of CDM and FDM cosmologies (as detailed in Section \ref{ss_sims}), exploring a range of axion masses spanning from $m=10^{-22}$ eV to $2\times 10^{-21}$ eV. Note that in this mass range, FDM is effectively ruled out as comprising $100$\% of the DM content \citep[as discussed in][]{Dome_2022}. However, in this study, the FDM-like modeling approach will allow us to elucidate trends associated with the axion mass $m$ across a variety of observables. These trends are expected to stay robust even if FDM constitutes only a subcomponent of the DM.\par 

In order to extract the maximum information from IM surveys, it is critical to reliably model the spatial distribution of HI. In our modeling approach, we largely follow the cell-based HI post-processing techniques from \cite{Villaescusa_2014, Villaescusa_2018_2, Carucci_2015, Carucci_2018, Diemer_2018, Diemer_2019}. We also harness the potential of machine learning (ML) and use normalizing flows, a generative ML model introduced by \cite{Agnelli_2010}, to capture intricate structures resulting from non-linear physics, facilitating the \textit{conditional} generation of HI maps with varying axion masses. Specifically, we use a modified version (see Sec. \ref{a_cond_nflows}) of the conditional normalizing flow framework HIGlow \citep{Friedman_2022} and showcase the efficiency of the model in sample generation. We assess the generated HI maps against external simulation data, utilizing metrics like the HI mass probability density function and HI power spectrum, validating the prowess of the model across a broad spectrum of mass scales and spatial dimensions. As proof of concept, we demonstrate the ability of HIGlow to interpolate the latent space of axion masses by predicting the peak flux in a synthetic cosmology, see Sec. \ref{s_mock_radio}. This affirms its efficacy in characterizing HI distributions for forthcoming parameter forecasting endeavors.\par 

The organization of the paper is as follows: In Sec. \ref{ss_sims}, we describe our CDM and FDM hydrodynamical simulations. We summarize our modeling of hydrogen phases and its link to 21 cm physics in Secs. \ref{ss_estimate_HI} and \ref{ss_21cm_intro}, respectively. In Sec. \ref{ss_higlow}, we evaluate the trained model. The post-reionization HI abundance, HI and 21 cm clustering (including the HI bias) and the HI column density distribution are investigated in Secs. \ref{ss_overall_HI}, \ref{ss_HI_power} and \ref{ss_HI_col}, respectively. For the first time, we analyze the cross-section of damped Lyman-$\alpha$ absorbers (DLAs) in FDM-like cosmologies (in Sec. \ref{ss_dla_cross_section}) and the imaging prospects of SKA-Low using mock radio maps, augmented by the trained normalizing flow (in Sec. \ref{s_mock_radio}). We conclude in Sec. \ref{s_conclusions}. HIGlow implementation choices are summarized in App. \ref{a_cond_nflows}.

\section{Theoretical Models and Post-Processing}
\label{s_num_methods}

\subsection{CDM and FDM-like Simulations}
\label{ss_sims}
We focus on axions generated via vacuum realignment assuming gravitational interactions do not re-thermalize axions. Simulating such FDM models is significantly more challenging than for CDM, as in the densest regions the wavelike matter oscillations can attain high frequencies $\omega \propto m^{-1}\lambda_{\text{dB}}^{-2}$, requiring very fine temporal resolution even for moderate spatial resolution \citep{Mocz_2019, May_2021}. To bypass the challenges of FDM simulations, here we employ FDM-like modeling described in \cite{Dome_2022}. In short, we impose a cutoff in the primordial power spectrum but evolve the system using only the CDM dynamics \citep[also see][]{Ni_2019}, providing a useful approximation for FDM and other DM scenarios incorporating a small-scale cutoff such as warm dark matter \citep[WDM,][]{Paduroiu_2022}. Following \cite{Dome_2022}, we call this proxy classical FDM (cFDM). For scales corresponding to wavenumbers $k \sim 0.16 - 80 \ h/\text{Mpc}$, which we explore here, the dynamical manifestation of FDM - the \textit{quantum pressure} \citep[fluid formulation,][]{Madelung_1927} - has only small impact on the growth of DM fluctuations. Specifically, the absolute fractional difference between growth rates in FDM vs cFDM is less than $5$\% for particle masses around $m\sim 10^{-22}$ eV and halo mass scales around $M \sim 4\times 10^9 \ M_{\odot}/h$ \citep[see e.g.][]{Corasaniti_2017}. As opposed to a superfluid, cFDM approximates FDM as a classical collisionless fluid, governed by the Vlasov-Poisson system of equations, but with FDM initial conditions. The exponential-like small-scale suppression in the primordial power spectrum, which we will often refer to as a cutoff, is modeled using the Boltzmann solver \scshape{AxionCamb}\normalfont \ \citep{Hlozek_2015}.\par 

We employ the \TNG \ galaxy formation module and run the simulations with the state-of-the-art code \mbox{\scshape{Arepo}} \normalfont described by \cite{Arepo_2010} and \cite{Weinberger_2020}. The hydrodynamical equations are solved on a moving Voronoi mesh using a finite volume technique. Various astrophysical processes such as metal-line cooling, star formation and feedback remain unresolved in the simulations and are approximated by subgrid models \citep{Pillepich_2017}. Gas above a density threshold of $n_{\text{H}} \sim 0.1$ cm${}^{-3}$ spawns star particles stochastically following the empirical Kennicutt-Schmidt relation and assuming a \cite{Chabrier_2003} initial mass function (IMF).\par

In our suite, we use cosmological volumes with a box side length of $L_{\text{box}} = 40 \ h^{-1}$Mpc. The box size balances the competing demands of high resolution of HI distributions and large volume (to obtain accurate statistical distributions). We vary the DM resolutions across $N=256^3$, $512^3$, and $1024^3$. For the bulk of this study, our primary focus is on high-resolution runs at $1024^3$ while we address resolution effects in our discussions and commentary throughout the text. In such full hydrodynamical runs, the imprints of cFDM are entangled with resolution effects that are due to baryonic physics \citep[e.g.][]{Vogelsberger_2013, Chua_2019}. Similar to \citet{Villaescusa_2018_2}, our findings related to HI distributions are not converged against resolution. Specifically, the HI and 21 cm power spectrum, HI abundance, HI column density distributions and DLA cross-sections exhibit resolution-dependent behavior. However, this does not undermine the validity of our results, as our work is conducted at the effective resolution level of \TNG100, on which the model parameters have been calibrated to accurately reproduce galaxy properties. Whilst our box size is smaller than that of \TNG100, our effective resolution is comparable, given the lower resolution setting of $1024^3$.\par 

Apart from CDM, we run cFDM simulations over a range of axion masses $m=10^{-22}, \ 7\times 10^{-22}, \ 2\times 10^{-21}$ eV. DM halos are identified using the friends-of-friends ({\fontfamily{cmtt}\selectfont FoF}) algorithm with a standard linking length of $b = 0.2 \times \text{(mean inter-particle separation)}$ \citep{Springel_2001_2}. As a minimum halo resolution, we require all halos to be composed of at least $200$ DM resolution elements. We adopt a Planck cosmology \citep{Planck_2015} with $\Omega_m = 0.3089$, $\Omega_{\Lambda} = 0.6911$, $h=H_0/100=0.6774$ and $\sigma_8 = 0.8159$. Initial conditions are set up at $z=127$, using $n_s = 0.9665$ for the primordial power spectrum of CDM and as input to \scshape{AxionCamb}\normalfont. We evolve the boxes to redshift $z=3.42$.\par 

\subsection{Estimating HI Fractions}
\label{ss_estimate_HI}
In our simulations, we aim to estimate the individual HI fraction in each gas cell. The \TNG \ model uses a modification of the two-phase interstellar medium (ISM) model \citep[][SH03]{Springel_2003}, which assumes star formation occurs only above a certain mass density threshold. Below the threshold, gas physics is determined by hydrodynamics assuming an ideal gas equation of state. Above the threshold, the gas is assumed to have two phases: cold star-forming clouds with a temperature of $1000$ K, and hot ionized gas. The SH03 model gives an effective mass-weighted specific energy to a gas cell by averaging over the cold and hot phases,
\begin{equation}
u_{\text{eff}} = (1-x)u_h + xu_c,
\end{equation}
where $u_{\text{eff}}$ is the effective specific energy, $x$ is the mass fraction of cold gas in a cell, and $u_h$ and $u_c$ are the energy per unit mass of the hot and cold gas, respectively. \TNG \ uses a modified effective temperature $u_{\text{eff,TNG}}$ given by $$u_{\text{eff,TNG}} = 0.3u_{\text{eff}} + 0.7u_4,$$
where $u_4$ is the energy per unit mass corresponding to a temperature of $10^4$ K. The HI fraction in each gas cell stored in the snapshots is based on this effective temperature. For star-forming cells, the effective temperature approach underestimates the HI fraction \citep{Villaescusa_2018_2}. One can get a better estimate via a direct calculation that does not use the effective temperature. We model the gas similarly as having two phases, and we assume the cold phase is completely comprised of HI, and the hot phase is fully ionized. The cold mass (i.e. HI) fraction $x$ \citep{Springel_2003} can then be expressed analytically as
\begin{equation}
x = 1 + \frac{1}{2y} - \sqrt{\frac{1}{y}+\frac{1}{4y^2}},
\end{equation}
where we define
\begin{equation}
y = \frac{t_* \Lambda_{\text{net}}(\rho , u_h)}{\rho (\beta u_{\text{SN}} - (1-\beta )u_c)}.
\end{equation}
Here $t_*$ is the star formation time scale,  $\Lambda_{\text{net}}$ is the net cooling rate, $\rho$ is the gas density, $u_{\text{SN}}$ is the specific energy corresponding to a supernova temperature of $T_{\text{SN}} = 5.73 \times 10^7 \text{ K}$, and $\beta$ is the mass fraction in massive stars. Note that we use the \TNG \ values of $t_* = 3.28 \text{ Gyr}$ and $\beta = 0.226$. Furthermore, we can convert between specific energies and temperatures using the equation 
\begin{equation}
T = \frac{(\gamma - 1) \mu}{k_B}u,
\end{equation}
where $\gamma$ is the adiabatic index, $k_B$ is the Boltzmann constant, and $\mu$ is the mean molecular weight given by 
\begin{equation}
\mu = \frac{4m_{\mathrm{p}}}{1+3X_H + 4X_H (1-x)}.
\end{equation}
Here, $m_{\mathrm{p}}$ is the mass of a proton and $X_H$ is the hydrogen mass fraction.\par
\begin{figure*}
\hspace{0.2cm}
\includegraphics[scale=0.90]{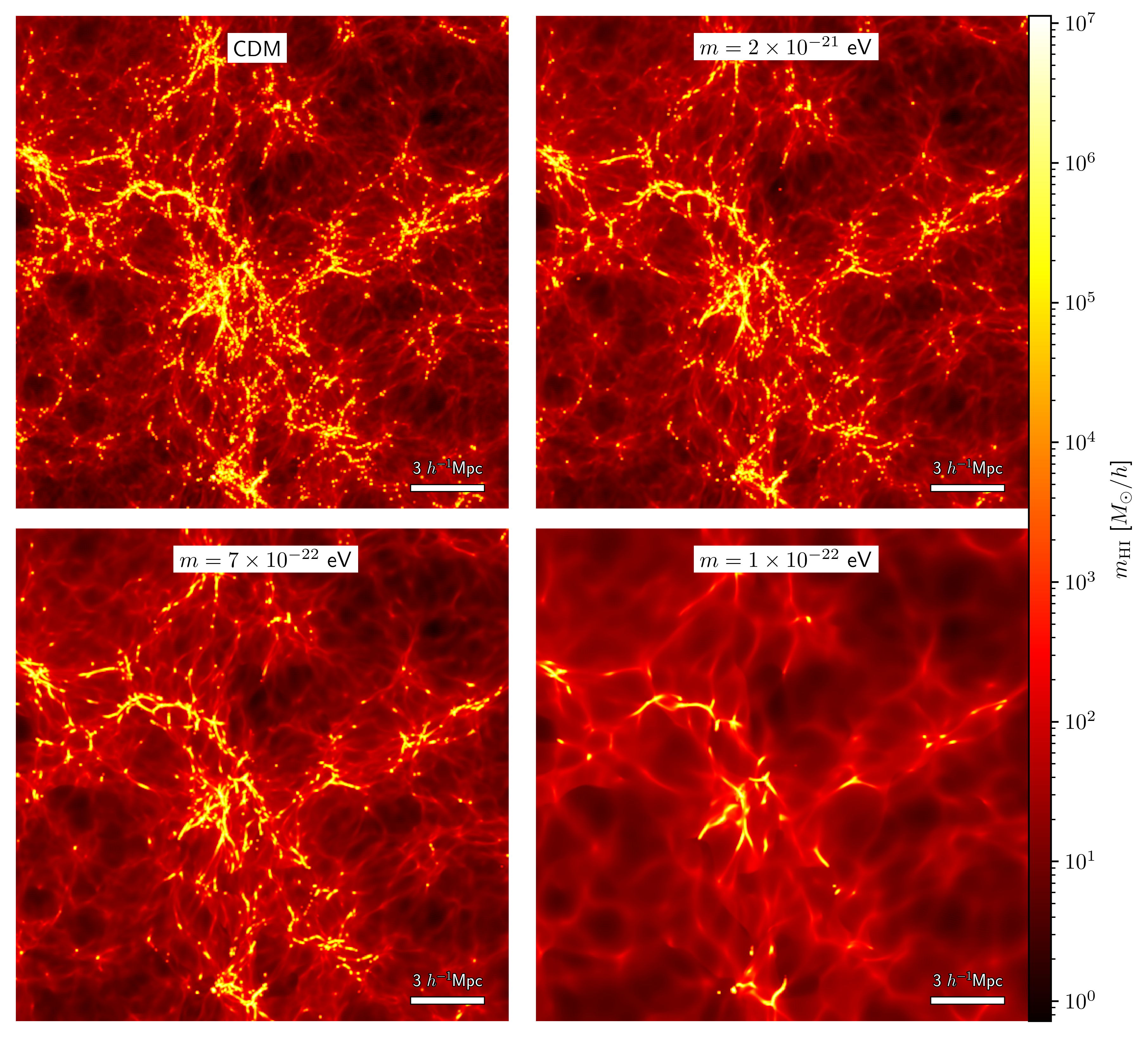}
\caption{HI mass maps in CDM and cFDM cosmologies at the post-reionization redshift of $z=4.94$. The width of the projected slice $\Delta L \approx 2.5 \ h^{-1}$Mpc corresponds to a channel width of $\Delta \nu = 272$ kHz. Redshift-space distortions (RSDs) are included by displacing Voronoi cells following Eq.~\eqref{e_rsd}.}
\label{f_HI_maps}
\end{figure*}
A further correction is applied to the HI fraction estimate by subtracting the molecular hydrogen fraction. We adopt the KMT model \citep{KMT_2009}, which expresses the mass fraction of molecular hydrogen to neutral hydrogen (both atomic and molecular) $f_{\text{H}_2}$ in star-forming cells as
\begin{equation}
f_{\text{H}_2}=\begin{cases}
1 - \cfrac{0.75s}{1+0.25s} & \text{if $s<2$,} \\
0 & \text{if $s \geq 2$,}
\end{cases}
\end{equation}
where $s$ is given by
\begin{equation}
s = \cfrac{\log(1 + 0.6\chi+0.01\chi^2)}{0.6\tau_c},
\end{equation}
and
\begin{equation}
\chi = 0.756(1+3.1Z^{0.365}), \ \ \tau_c = \Sigma \sigma_d / \mu_{\text{H}}.
\end{equation}
In the above equations, $Z$ is the gas metallicity (in units of solar metallicity), $\sigma_d = Z \times 10^{-21} \text{ cm}^2 $ is the cross-section of dust, $\mu_{\text{H}}=2.3 \times 10^{-24} \text{ g}$ is the mean mass per hydrogen nucleus, and $\Sigma = \rho R$ is the surface density of the gas, where $\rho$ is the gas density and the volume $V$ of the cell can be translated to an effective radius $R=(3V/4\pi)^{1/3}$.\par

Fig. \ref{f_HI_maps} shows 2D HI mass projections of post-processed $\Lambda$CDM and cFDM simulation data. We choose a frequency channel \mbox{$[\nu_{0}-\Delta/2, \nu_{0}+\Delta/2]$} of width $\Delta \nu = 272$ kHz around the central frequency $\nu_0 = \nu_{21}/(1+z)$, where $\nu_{21} = 1420.406$ MHz is the rest frequency of the 21 cm line. The channel width corresponds to a comoving width of
\begin{equation}
\Delta L = r_{\nu_{21}-\Delta \nu/2} - r_{\nu_{21}+\Delta \nu/2} \approx 2.5 \ h^{-1}\text{Mpc},
\end{equation}
where $r_{\nu}$ is the comoving distance to redshift \begin{equation}
z=\nu_{21}/\nu - 1,
\label{e_nu_z_rel}
\end{equation}
for the observational frequency $\nu$. We take the same slice of width $\Delta L$ along the $x$-direction from the four hydrodynamic snapshots at $z=4.94$, PCS-paint the HI particle data onto a 3D grid \citep{Hand_2018}, and project along the $x$-direction. We see how small-scale structure is visibly suppressed as the axion mass $m$ is reduced from infinity (CDM, top left) down to $m = 10^{-22}$ eV (bottom right).

\subsection{21 cm Cosmology}
\label{ss_21cm_intro}
We focus on the post-reionization era of $z<5.3$, in which most of the hydrogen has been ionized and the volume-averaged intergalactic medium (IGM) HI fraction has been reduced to $\bar{x}_{\mathrm{HI}} \approx 10^{-4}$ according to Lyman-$\alpha$ (Ly$\alpha$) and Lyman-$\beta$ (Ly$\beta$) effective optical depth measurements \citep{Yang_2020, Bosman_2021}. The surviving HI resides in dense regions of matter that are able to self-shield against ionizing background radiation. While most of the HI is found within halos, typically of mass $M_{\mathrm{h}} \sim 10^{10}-10^{13} \ M_{\odot}$, at $z=5$ around $12$\% of the HI is found outside of halos \citep{Villaescusa_2018_2}.\par 

The ratio between the number of HI atoms in the ground state and the excited spin state depends on the spin temperature $T_s$, and is given by 
\begin{equation*}
\frac{n_1}{n_0} = 3\exp \left(-\frac{h\nu_{21}}{k_BT_s}\right),
\end{equation*}
where $n_0$ and $n_1$ are the number of ground and excited spin states respectively and $h$ is Planck’s constant.\par
\begin{figure*}
\hspace{0.2cm}
\includegraphics[scale=0.90]{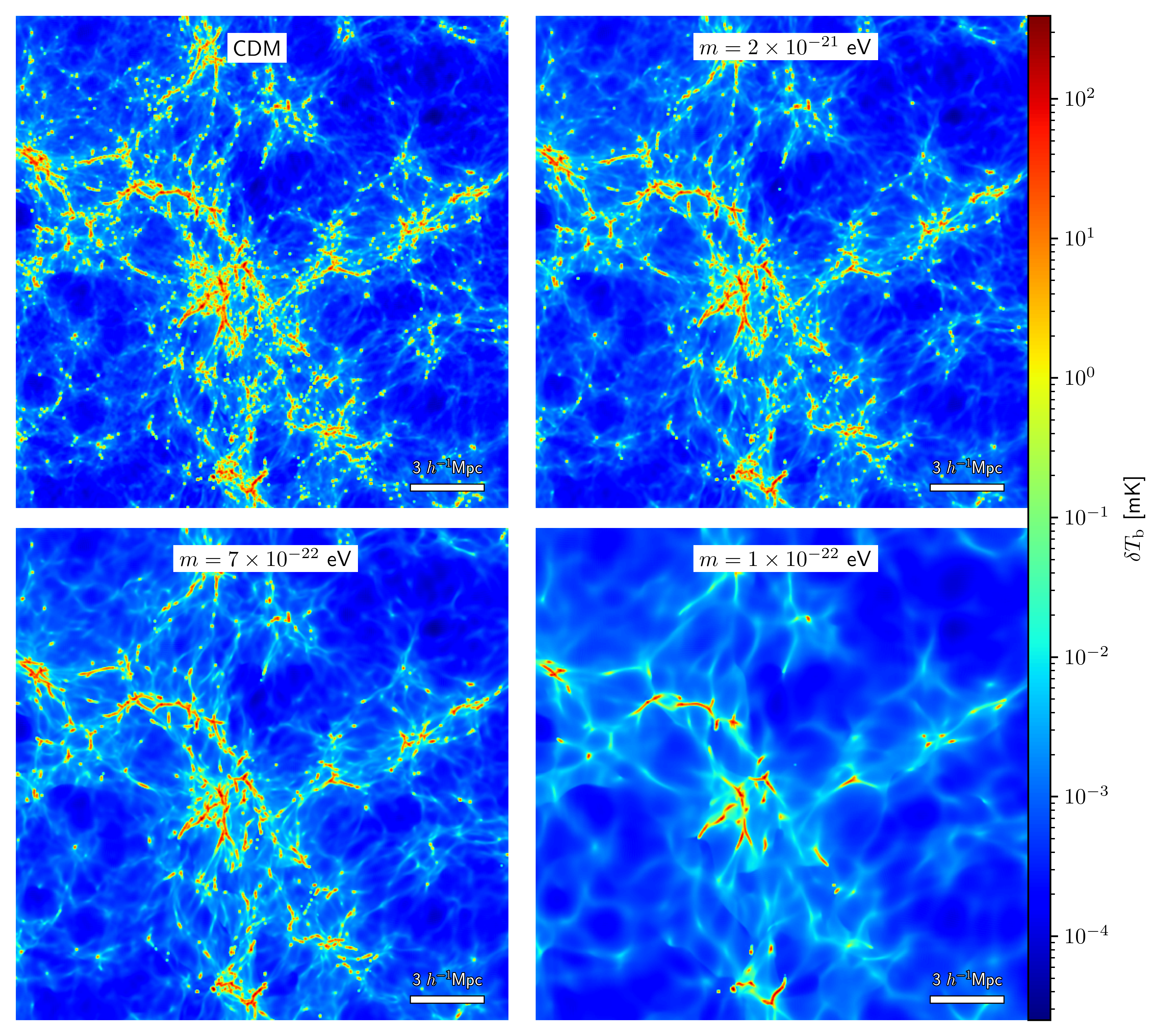}
\caption{21 cm maps in various cosmologies at $z=4.94$. The comoving projections are over the same width as in Fig. \ref{f_HI_maps}. The highest 21 cm brightness temperatures $\delta T_{\mathrm{b}} \approx 350$ mK are found in DLAs which can self-shield against external UV radiation.}
\label{f_21_maps}
\end{figure*}
The differential brightness temperature $T_b$ describes the difference between the CMB temperature $T_\gamma$ and the spin temperature $T_s$ and is given by
\begin{equation*}
\delta T_b(\hat{\mathbf{r}},\nu) = \left[1-e^{-\tau_{21}(\hat{\mathbf{r}},z)}\right]\frac{T_s(\hat{\mathbf{r}},z)-T_{\gamma}(z)}{1+z},
\end{equation*}
where $\hat{\mathbf{r}}$ is a unit vector that is centred on the observer
and points in the direction of observation on the sky. The observational frequency $\nu$ is related to the redshift $z$ by the standard relation, Eq. \eqref{e_nu_z_rel}. We have also defined the optical depth $\tau_{21}$ across the 21 cm line at redshift $z$ \citep[]{Pritchard_2012, Liu_2020},
\begin{equation*}
\tau_{21}(\hat{\mathbf{r}},z) = \frac{3h c^3A_{10}}{32\pi k_B\nu_{21}^2}\frac{x_{\mathrm{HI}}n_{\mathrm{H}}}{(1+z)(dv_{||}/dr_{||})T_s},
\end{equation*}
where $n_{\mathrm{H}}$ is the hydrogen density, $dv_{||}/dr_{||}$ the gradient of the proper velocity along the line of sight, $c$ the speed of light and $A_{10}$ the Einstein spontaneous emission coefficient for the hyperfine transition.\par

Although observationally the HI can be part of either cold ($T\lesssim 100$ K) or warm ($T\gtrsim 5000$ K) neutral medium, both temperatures are warmer than the CMB temperature (which is $19.1$ K at $z=6$) in the post-reionization era due to heating by sources of radiation such as stars and galaxies. In \TNG \ subgrid modeling, low-temperature cooling channels are not included, leading to a putative cooling floor of $10^4$ K. We indeed find that very few gas cells have temperatures below $\approx 3\times 10^3$ K in the post-reionization era. Due to \Lya \ coupling \citep[see e.g.][]{Furlanetto_2006}, the spin temperature of HI is thus expected to be higher than the CMB temperature, and we expect to see the 21 cm line in emission. Observationally, the fiducial spin temperature of $T_s = 100$ K is often assumed for DLAs due to a lack of specific measurements, which corresponds to assuming that all of the absorbing gas is in the thermally stable cold neutral medium phase \citep{Murray_2018}. In case the gas-phase metallicity, dust abundance and background UV field can be estimated, $T_s$ can be inferred, but reported values rarely fall below $T_s = 50$ K \citep{Zwaan_2015}. It is only in very rare environments where the \Lya \ radiation field is extremely weak and collisional processes are not effective, that the spin temperature might approach or even fall below the CMB temperature, challenging our assumption $T_\gamma/T_s \approx 0 $.\par 

Even though certain DLAs at low redshift $z<1$ can have peak optical depths (as inferred from the 21 cm line center) of $\tau_{21} \approx 0.5$ or beyond \citep{Curran_2017, Sadler_2020}, it is customary to Taylor-expand the brightness temperature,
\begin{equation}
\delta T_b(\hat{\mathbf{r}},\nu) = \frac{3h c^3A_{10}}{32\pi k_B\nu_{21}^2} \left[\frac{x_{\mathrm{HI}}n_{\mathrm{H}}}{(1+z)^2(\mathrm{d}v_{||}/\mathrm{d}r_{||})}\right].
\label{e_bt}
\end{equation}
In fact, the assumption $T_\gamma/T_s \approx 0 $ compels us to Taylor-expand to first order, i.e. these two simplifications that are very common in the literature go hand in hand. We expect the peak values of the brightness temperature field $\delta T_b$ as well as the peak values of the specific intensity field (see Sec. \ref{s_mock_radio}) to be slightly overestimated. However, we choose to maintain consistency with the literature since improving the modeling would necessitate a better understanding of the spin temperature $T_s$ in optically thick regions.\par

Eq. \eqref{e_bt} makes clear that the observation of the 21 cm signal can in turn be used as a tracer of matter or as an indirect probe of other properties of our universe, such as its ionization state or temperature. It is also sensitive to cosmology, since the distribution of hydrogen is based on the large-scale distribution of matter, as is the $\mathrm{d}v_{||}/\mathrm{d}r_{||}$ velocity term, since it includes peculiar velocities in addition to the Hubble flow. Specifically, we make use of the plane-parallel approximation to displace the positions of Voronoi cells from real space $(\mathbf{r})$ to redshift space $(\mathbf{s})$ through
\begin{equation}
\mathbf{s} = \mathbf{r}+\frac{1+z}{H(z)}\mathbf{v}_{||}(\mathbf{r}),
\label{e_rsd}
\end{equation}
where $H(z)=H_0\sqrt{\Omega_{\text{r,0}}(1+z)^4+\Omega_{\text{m,0}}(1+z)^3+\Omega_{\Lambda,0}}$ is the familiar Hubble parameter and $\mathbf{v}_{||}(\mathbf{r})$ is the peculiar velocity of the cell along the line of sight. While it may seem that it is difficult to probe any of the effects contributing to $T_b$ cleanly, different phenomena tend to dominate at different redshifts.\par

Note that in our \TNG \ modeling approach, there is a simplifying assumption in the implementation of reionization. The UV background is turned on gradually around $z \approx 10$ as a homogeneous, isotropic radiation field, instead of implementing more realistic patchy reionization histories \citep{Byrohl_2021, Molaro_2022, Bird_2023, Puchwein_2023} which significantly affect the abundance, distribution, and properties of low column density HI absorbers, changing the shape, distribution, and strength of absorption lines within the \Lya \ forest. Inhomogeneous reionization not only modifies the post-reionization 21 cm power spectrum and the values for the cosmological parameters inferred from IM experiments \citep{Long_2023_2} but also the abundance and metallicity distribution of DLAs \citep{Hassan_2020_2}. We ignore such effects in this work since in the post-reionization era, the assumption of a uniform UV background leads to good agreement with observations of the mean \Lya \ flux, the HI abundance and the distribution of intermediate and high HI column densities \citep[see Sec. \ref{s_HI_results} and e.g.][]{Villaescusa_2018_2}.\par 

In Fig. \ref{f_21_maps}, we show 21 cm maps in CDM and cFDM cosmologies at $z=4.94$ using this approximation. While the brightness temperature $\delta T_{\text{b}}$ is strictly positive throughout the maps, its magnitude spans a large range of over $7$ orders of magnitude. The maximum brightness temperature reach $\delta T_{b,\text{max}} \approx 350$ mK and are typically associated with DLAs (see Sec. \ref{ss_HI_col}). Note that $\delta T_{b,\text{max}}$ is highly dependent on numerical resolution. For instance, peak values at $z=4.94$ for the $512^3$ runs are notably reduced to $\delta T_{b,\text{max}}\approx 95$ mK.\par

\section{Post-Reionization HI Distributions}
\label{s_HI_results}

\subsection{HIGlow Modeling of HI}
\label{ss_higlow}
\begin{figure}
\hspace{-0.3cm}
\includegraphics[width=0.5\textwidth]{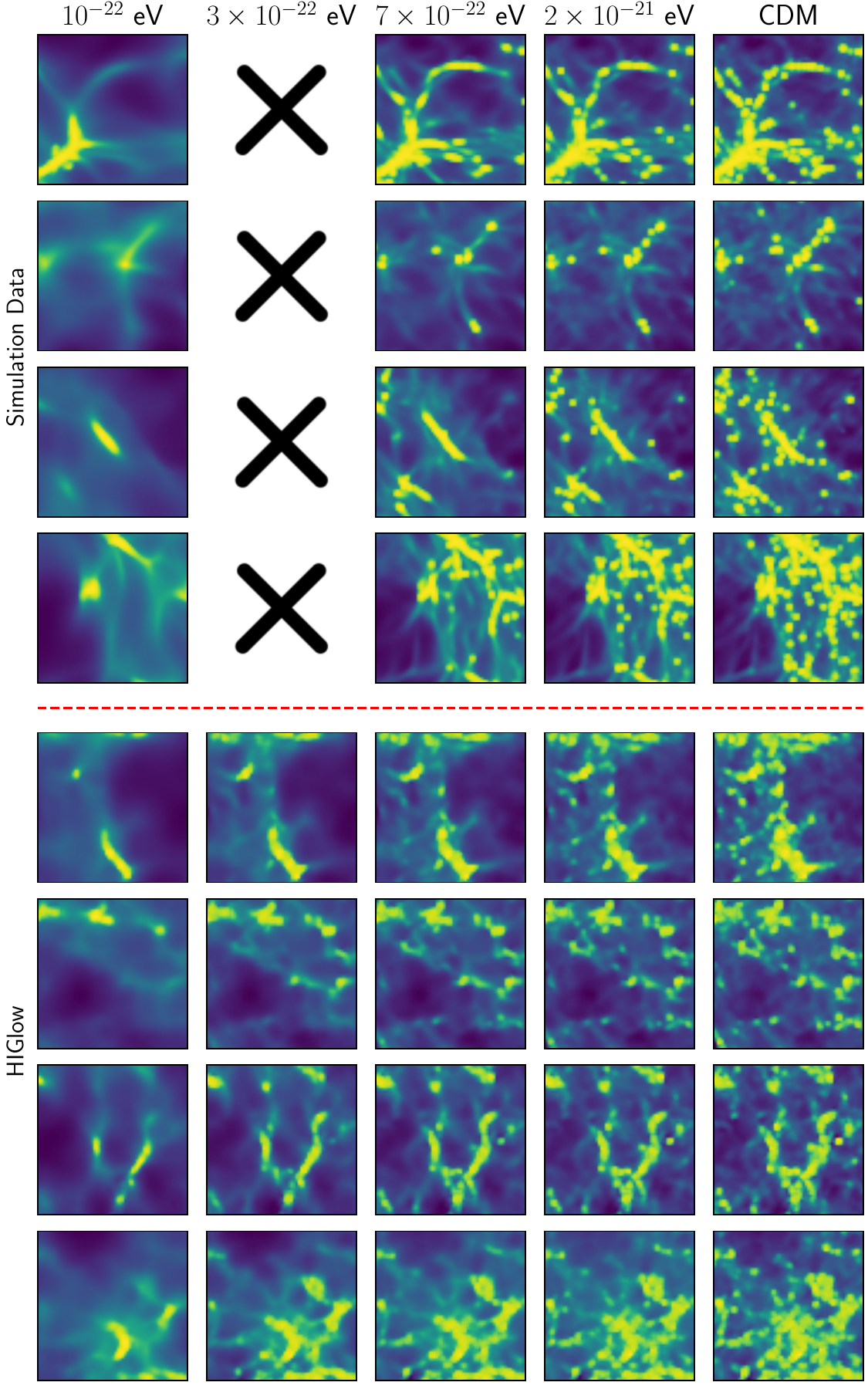}
\caption{Simulated data training samples (top four rows) and generated samples with HIGlow (bottom four rows) in the various cosmologies at $z=4.94$. The $64^2$ maps are $2.5 \ h^{-1}$Mpc a side. We add HIGlow samples for a conditional axion mass $m = 3\times 10^{-22}$ eV. This case does not have corresponding simulation data, illustrated by crosses. The axion mass $m$ decreases from right to left, with increased small-scale suppression.}
\label{f_samples}
\end{figure}
To model HI distributions in the post-reionization era, we utilize normalizing flows, a class of generative models which are hailed for their expressiveness in modeling complex distributions with exact likelihoods and for providing an invertible mapping between a simple base distribution and the target distribution, enabling easy sampling and generation of realistic data. Details of our HIGlow implementation are given in App. \ref{a_cond_nflows}.\par 

The top four rows of Fig. \ref{f_samples} show $64\times 64$ HI training maps across CDM and cFDM cosmologies at $z=4.94$. The bottom four rows show generated HIGlow HI samples with identical seeds across the cosmologies. The samples visually have very similar features compared to the input training data. The model has learned the effect of changing the axion mass $m$ on the HI maps, and the characteristic suppression of small-scale structure at lower axion mass is apparent. We also show HIGlow samples for a synthetic cosmology corresponding to a new axion mass $m=3\times 10^{-22}$ eV. The generated samples display subtle small-scale features that occupy an intermediate position between those of the $m=10^{-22}$ eV and $m=7\times 10^{-22}$ eV samples, aligning with our expectations. We have conducted comprehensive validation tests, the results of which are outlined in the following section.

\begin{figure}
\hspace{-0.3cm}
\includegraphics[width=0.5\textwidth]{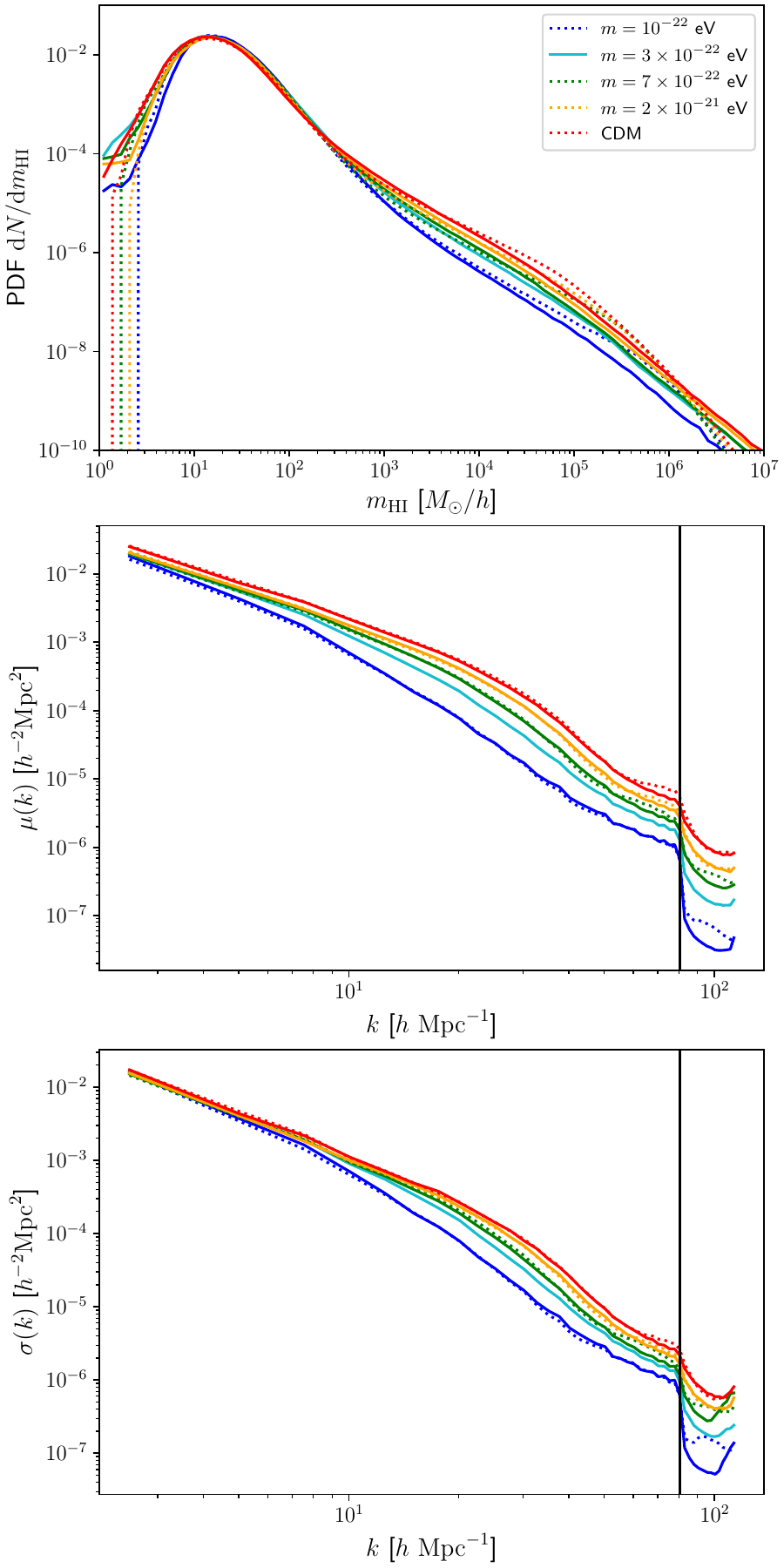}
\caption{Statistical properties of simulation data (dotted lines) and HIGlow-generated HI samples (solid lines) at $z=4.94$ in CDM and cFDM cosmologies. We show HI mass PDFs (top panel), the median of 2D power spectra among $1000$ random samples (middle panel) and their standard deviation (bottom panel). In cyan, we add the results for a conditional axion mass of $m = 3\times 10^{-22}$ eV (see Fig. \ref{f_samples}). This case does not have corresponding simulation data. The Nyquist frequency $k_{\text{Ny}} = \pi N^{1/3} / L_{\text{box}}$ is shown as a vertical line.}
\label{f_new_ma}
\end{figure}
To ensure that the statistics of the generated HI maps follow the statistics of the simulation data, we use two validation metrics: HI mass PDFs and HI power spectra. Fig. \ref{f_new_ma} shows the HI mass PDFs as well as the mean power spectrum and the standard deviation for $1000$ simulation data subcube projections (dotted lines) and the same number of HIGlow-generated samples (solid lines). The HI mass PDFs of the generated samples closely follow the target distribution, especially for HI masses below $m_{\text{HI}} \approx 10^{5} \ M_{\odot}/h$. In the range $m_{\text{HI}} = 10^5-10^7 \ M_{\odot}/h$, HIGlow captures the distribution less accurately, which is a result of the rarity of such high-mass pixels. For instance, $m_{\text{HI}} = 10^6 \ M_{\odot}/h$ pixels are $6-7$ orders of magnitude less likely than pixels with $m_{\text{HI,peak}} \approx 20 \ M_{\odot}/h$. The PDF bump around this peak value broadly corresponds to the \Lya \ forest in the post-reionization era \citep[also see][]{Zamudio_2019}.\par 

Note that before training HIGlow, the data undergoes sigmoid normalization as per Eq. \eqref{e_norm}. The samples depicted in Fig. \ref{f_samples} are presented in sigmoid space, and similarly, as are the power spectra showcased in Fig. \ref{f_new_ma}. This choice enhances the clarity of trends concerning the axion mass $m$. We find that the target power spectra are reproduced with high fidelity, in particular the mean and standard deviation. The sigmoid-normalized power spectra exhibit the opposite trend with axion mass compared to the 3D non-sigmoid-normalized power spectra \citep[see Sec. \ref{ss_HI_power} and e.g.][]{Carucci_2015}. We have checked and find that while the projection effect reduces the relative difference between cosmologies as reflected in 2D vs 3D power spectra, the sigmoid normalization \textit{reverses} the trend.\par 

In the evaluation of HIGlow, a conditional generative model, testing its capability to generate diverse samples spanning the entire latent space of axion masses is crucial. We test this functionality by choosing a new axion mass of $m = 3\times 10^{-22}$ eV, unexplored by the simulations, and generating $1000$ random images with HIGlow (see Fig. \ref{f_samples}). Note that (unlike for CDM and the other three axion masses) we only show the HIGlow generated data for $m = 3\times 10^{-22}$ eV, as the corresponding simulation data does not exist for this mass. The statistics of the synthetic cosmology are shown in Fig. \ref{f_new_ma}, exhibiting a striking resemblance to the anticipated outcome. Falling in between the curves corresponding to $m = 10^{-22}$ eV and $m = 7 \times 10^{-22}$ eV, the generated distribution effectively showcases the interpolation prowess in latent space. While a direct comparison to actual simulations at $m = 3 \times 10^{-22}$ eV would be needed for a more definitive test, the achieved success can be attributed to the monotonic albeit nonlinear influence of the axion mass $m$ on the distribution of HI.

\subsection{Overall HI Abundance}
\label{ss_overall_HI}
Here we investigate the overall HI abundance $\Omega_{\text{HI}} = \bar{\rho}_{\text{HI}}(z)/\rho_{\mathrm{crit}}^0$ relative to the critical density of the Universe today $\rho_{\mathrm{crit}}^0$, with $\bar{\rho}_{\text{HI}}(z)$ being the mean HI density at redshift $z$. Direct measurements from HIPASS \citep{Barnes_2001} and ALFALFA \citep{Jones_2018, Oman_2022} have been used to detect individual extra-galactic objects, allowing constraints on the HI mass function and abundance of low redshift ($z\approx 0$) HI at around $\Omega_{\mathrm{HI}} \sim (3.9 \pm 0.6) \times 10^{-4}$. At slightly higher redshift, cross-correlations with optical tracers (e.g. DEEP2 or WiggleZ) lead to estimates of the HI abundance at $z\sim 0.8$ to be $\Omega_{\mathrm{HI}}b_{\mathrm{HI}} \sim 6.2^{+2.3}_{-1.5} \times 10^{-4}$ \citep[]{Switzer_2013}. More indirect measurements of DLAs allow us to constrain the HI abundance up to $z\sim 5$. Therefore, in the post-reionization era, we have a well-defined target for the amplitude of the 21 cm signal, which is proportional to the square of the HI abundance, $P_{21} \propto \Omega_{\text{HI}}^2$.\par

\begin{figure}
\hspace{-0.6cm}
\includegraphics[scale=0.58]{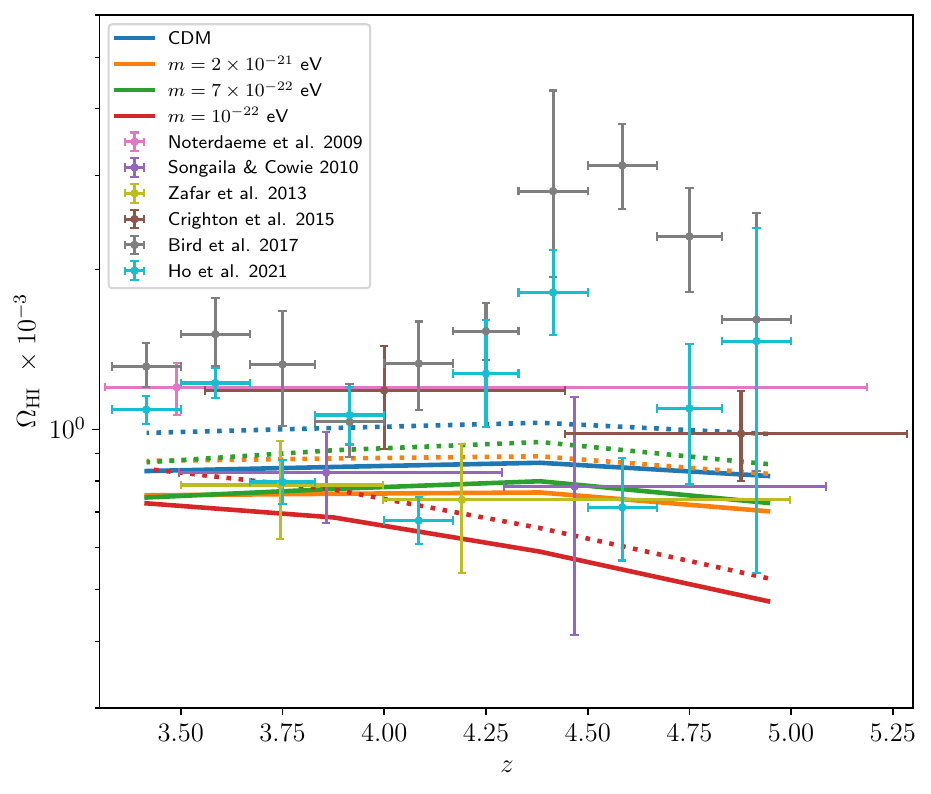}
\caption{Comparison of mock and observed cosmic HI density estimates. The abundance parameter $\Omega_{\text{HI}}=\bar{\rho}_{\text{HI}}(z)/\rho_c^0$ from simulations in the redshift range $z=3.42-4.94$ for various cosmologies is shown by dotted lines (in real space) and solid lines (in redshift space), obtained using the procedure outlined in Sec. \ref{ss_estimate_HI}. Observational measurements are displayed with $1\sigma$ error bars: \protect\cite{Songaila_2010} provide DLA measurements from Keck data; \protect\cite{Zafar_2013} quote combined DLA and sub-DLA measurements from ESO UVES; \protect\cite{Crighton_2015} quote results from a Gemini GMOS study of DLAs; \protect\cite{Noterdaeme_2009}, \protect\cite{Bird_2017} and \protect\cite{Ho_2021} are DLA analyses using SDSS DR7, DR12 and DR16Q, respectively.}
\label{f_OmegaHI}
\end{figure}

We show the value of $\Omega_{\text{HI}}(z)$ in the four simulated cosmologies in Fig. \ref{f_OmegaHI} in the redshift range $z=3.42-4.94$. HI abundances estimated in redshift space are typically lower than in real space since the `squashing' effect from coherent large-scale flows is subdominant compared to the Fingers of God effect driven by virial motions (see Sec. \ref{ss_HI_power}). To compare our predictions to observations, we first note that in a typical DLA survey, HI column densities are estimated from absorption spectra of quasars. The HI frequency distribution or column density distribution function (CDDF, also see Sec. \ref{ss_HI_col}) is typically written as
\begin{equation}
f_{\text{HI}}(N_{\text{HI}}, X) = \frac{\mathrm{d}^2N(N_{\text{HI}})}{\mathrm{d}N_{\text{HI}}\mathrm{d}X},
\label{e_HI_CDDF}
\end{equation}
where $N$ is the number of lines of sight with column densities between $N_{\text{HI}}$ and $N_{\text{HI}}+\mathrm{N_{\text{HI}}}$, and
\begin{equation}
\mathrm{d}X = \frac{H_0}{H(z)}(1+z)^2\mathrm{d}z
\end{equation}
is the so-called absorption distance. The absorption distance has the property that absorbers with non-evolving number density $n_a$ and cross-section $\sigma_a$ (both in proper length units) will produce a constant number of absorption lines per unit absorption distance, i.e. $\mathrm{d}N/\mathrm{d}X = \text{const}$.\par 
\begin{figure*}
\includegraphics[width=\textwidth]{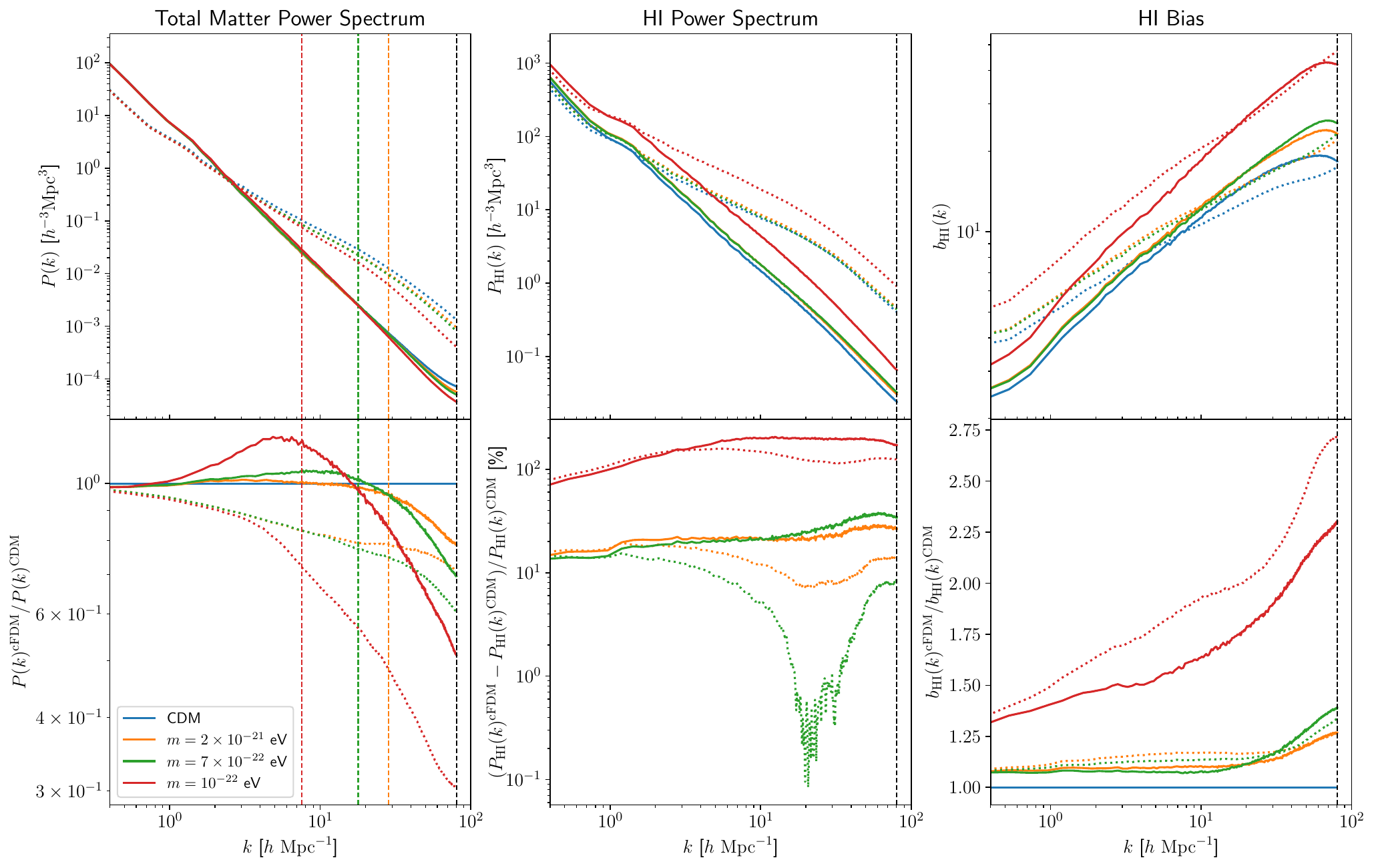}
\caption{Total matter (left panel), HI power spectrum (middle panel) and HI bias (right panel) in CDM and cFDM cosmologies at $z=4.94$. Each panel shows results in real space (dotted lines) and redshift space (solid lines). The vertical dashed lines mark the cFDM half-mode scale $k_{1/2}$ (colored) as per \protect\cite{Marsh_2016} and the Nyquist frequency $k_{\text{Ny}} = \pi N^{1/3} / L_{\text{box}}$ (in black). In the bottom row, we present the ratio of results between cFDM and CDM, with the first and third panels displaying the absolute ratio and the second panel showing the relative difference in percent. Both HI clustering and HI bias $b_{\text{HI}}^2(k) = P_{\text{HI}}(k)/P_{\text{tot}}(k)$ increase for decreasing axion mass $m$.}
\label{f_HI_psp_bias}
\end{figure*}
The distribution $f_{\text{HI}}(N_{\text{HI}}, X)$ spans ten orders of magnitude from around $N_{\text{HI}}=10^{12}$ cm$^{-2}$, below which \Lya \ absorbers produce absorption lines too weak to be recognized, to $10^{22}$ cm$^{-2}$ above which absorbers are very rare. Quasar absorbers with HI column densities below $\simeq 10^{17.3}$ cm$^{-2}$ are called the \Lya \ forest, and physically represent the low-density and highly ionized gas that resides in the IGM. Systems with column densities above $N_{\text{HI,DLA}}=10^{20.3}$ cm$^{-2}$ are called DLAs, which typically correspond to extra-galactic regions and are self-shielded against external radiation. Systems with column densities in between these two regimes, are called Lyman Limit Systems (LLSs) in the range $10^{17.3}-10^{19.0}$ cm$^{-2}$ and sub-DLAs in the range $10^{19.0}-10^{20.3}$ cm$^{-2}$. The DLAs account for most of the HI, but it is the \Lya \ forest that accounts for most of the gas (neutral + ionized).\par

The total gas density can be inferred from the HI CDDF via
\begin{equation}
\Omega_{\mathrm{g}}^{\mathrm{DLA}}(X)\mathrm{d}X = \frac{\mu m_{\mathrm{p}}H_0}{c\rho_{\mathrm{crit}}^0}\int_{10^{20.3}}^{\infty}N_{\mathrm{HI}}f_{\mathrm{HI}}(N_{\mathrm{HI}}, X)\mathrm{d}X,
\label{e_total_gas_from_cddf}
\end{equation}
where $\mu = 1.3$ is the mean molecular mass of the gas. In the discrete limit, $\Omega_{\mathrm{g}}^{\mathrm{DLA}}$ is given by
\begin{equation}
\Omega_{\mathrm{g}}^{\mathrm{DLA}} = \frac{\mu m_{\mathrm{p}}H_0}{c\rho_{\mathrm{crit}}^0}\frac{\sum N_{\mathrm{HI}}}{\Delta X},
\end{equation}
where the sum is calculated for systems with $\log_{10}N_{\mathrm{HI}}\geq 20.3$ along lines of sight with a total pathlength of $\Delta X$. As noted in \citep{Zafar_2013, Crighton_2015}, we can convert the gas mass in DLAs $\Omega_{\mathrm{g}}^{\mathrm{DLA}}$ to the neutral hydrogen mass via
\begin{equation}
\Omega_{\mathrm{HI}} = \delta_{\mathrm{HI}}\Omega_{\mathrm{g}}^{\mathrm{DLA}}/\mu,
\end{equation}
where $\mu$ accounts for the mass of helium and $\delta_{\mathrm{HI}}=1.2$ estimates the contribution from systems below the DLA threshold of $10^{20.3} $ cm$^{-2}$. In Fig. \ref{f_OmegaHI}, we add a compilation of measurements, each accompanied by error bars, sourced from multiple studies \citep{Noterdaeme_2009, Songaila_2010, Zafar_2013, Crighton_2015, Bird_2017, Ho_2021}. Note that all the estimates have been corrected for helium content and cosmological factors. Specifically, both the Hubble constant $H_0$ and the absorption distance $\Delta X$ in Eq. \eqref{e_total_gas_from_cddf} have been accounted for.\par 

Within the error bars, the agreement between the results from simulations and observations is good. Hydrodynamic simulations like ours using the \TNG \ galaxy formation module typically agree better with observations than semi-analytic models \citep{Lagos_2014} and are comparable to the agreement found using the EAGLE simulations \citep{Rahmati_2015}. The trend of decreasing HI abundance towards smaller axion mass $m$ agrees well with the WDM results of \cite{Carucci_2015}\footnote{The thermal relic WDM masses explored in \cite{Carucci_2015} correspond to axion masses of $m\in [1.6\times 10^{-22},5.5\times 10^{-21}]$ eV.}.\par

We note that similar to \cite{Villaescusa_2018}, the overall HI abundance depends on resolution. The lower-resolution $N=512^3$ simulations contain $\sim 10$\% less HI in the explored redshift interval $z=3.42-4.94$ than the higher-resolution $N=1024^3$ runs shown in Fig. \ref{f_OmegaHI}. HI can reside in small halos of mass $M_h \sim 10^9 \ M_{\odot}/h$, few of which are resolved by the $N=512^3$ runs. For halo-based HI estimations such as those based on the halo model \citep{Villaescusa_2018_2}, the trend with changing resolution is most intuitive, but it is also reproduced in particle-based models as we find.

\subsection{HI Power Spectrum and Bias}
\label{ss_HI_power}
In Fig. \ref{f_HI_psp_bias}, we show the total matter and HI power spectrum estimated using a third-order piecewise cubic spline (PCS) mass-assignment scheme \citep[MAS,][]{Hand_2018} and subsequent Fourier transformation,
\begin{equation}
P_{X} = \frac{1}{N_{\text{modes}}}\sum_{\mathbf{k}\in k}\delta_{X}(\mathbf{k})\delta_{X}^{\ast}(\mathbf{k}),
\end{equation} 
with $X = \lbrace \text{tot}, \text{HI}\rbrace$ and $N_{\text{modes}}$ being the number of modes lying in the spherical shell $k$ of width $\delta k = 2\pi / L_{\text{box}}$. When calculating the HI power spectrum, we do not account for the fact that the actual density profile of the gas particles is not given by a uniform cube but described by the SPH kernel. A reliable amplitude mode correction implementation is beyond the scope of this paper \citep[also see][]{Villaescusa_2014}. However, we compensate for the MAS window function to improve the power spectrum fidelity on scales close to the Nyquist frequency $k_{\text{Ny}} = \pi N^{1/3} / L_{\text{box}}$, where $N^{1/3}$ is the number of cells per box side.\par
\begin{figure}
\hspace{-0.3cm}
\includegraphics[width=0.5\textwidth]{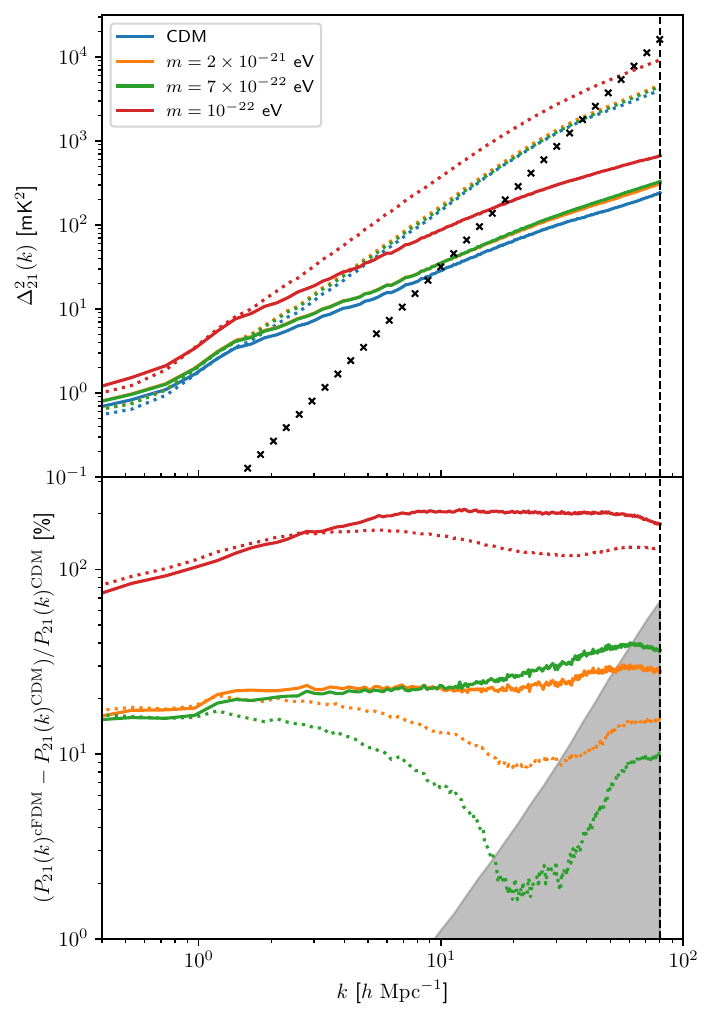}
\caption{21 cm power spectrum in CDM and cFDM cosmologies at $z=4.94$. We show results for the `dimensionless' power spectrum $\Delta^2_{21}(k)=k^3P_{21}(k)/(2\pi^2)$, in both real space (dotted lines) and redshift space (solid lines). Black crosses denote the expected SKA1-Low $1\sigma$ system noise $\Delta^2_{\text{noise}}$ estimated using \mbox{\scshape{21cmSense}} \normalfont \citep{Pober_2013,Pober_2014} for $1080$ hours of observation in a $8$ MHz bandwidth with $290$ frequency channels as per Table \ref{t_skalow_params}, for an optimistic foreground wedge model. The vertical dashed line denotes the Nyquist frequency $k_{\text{Ny}} = \pi N^{1/3} / L_{\text{box}}$ (in black). In the bottom panel, we present the relative difference between cFDM and CDM, identical to the relative difference in the HI power spectrum, see Fig. \ref{f_HI_psp_bias}. The normalized error $\Delta^2_{\text{noise}}/\Delta^2_{21,\text{CDM}}$ on the CDM 21 cm power spectrum for the same instrument parameters in SKA1-Low is shown by a shaded region. 21 cm clustering increases for decreasing axion mass $m$.}
\label{f_21_psp}
\end{figure}
In cFDM cosmologies, the total matter power spectrum in Fig. \ref{f_HI_psp_bias} follows the well-known trend of small-scale suppression for wavenumbers above $k_{1/2}$ \citep{Marsh_2016}. The effect of peculiar velocities, also called redshift-space distortions (RSDs), can be clearly discerned. On large scales, the clustering of matter in redshift space is enhanced due to the Kaiser effect \citep{Kaiser_1987}. On small scales, the peculiar velocities, particularly inside halos, give rise to the Fingers of God, suppressing the amplitude of the total matter power spectrum \citep{Tegmark_2004}.\par

HI is more clustered than the total matter since the UV background ionizes hydrogen in environments that are not dense enough to self-shield. As opposed to the suppression beyond $k_{1/2}$ in the total matter clustering, the HI power spectrum does not exhibit a small-scale cut-off since most of the HI is trapped in intermediate- and high-mass halos. Contrary to a naive expectation, HI clustering increases with decreasing axion mass $m$. Since most of the HI is locked inside DM halos in the post-reionization era, the higher halo bias in cFDM cosmologies compared to CDM \citep[see e.g.][]{Dunstan_2011} leads to an increased HI clustering. This is in agreement with \cite{Carucci_2015} who showed that the amount of HI per total mass in a given DM halo mass bin is very similar between CDM and WDM cosmologies for particle-based HI modeling approaches. As a result, the spatial distribution of HI is more biased in the cFDM models than in CDM, with the bias increasing with decreasing axion mass $m$. In the extreme cFDM model with $m=10^{-22}$ eV, deviations from CDM in the HI power spectrum can reach more than $300$\% at $z=4.94$  \citep[in agreement with WDM results from][]{Carucci_2015}.\par

The HI bias
\begin{equation}
b_{\text{HI}}(k) = \sqrt{\frac{P_{\text{HI}}(k)}{P_{\text{tot}}(k)}}
\label{e_HI_bias}
\end{equation}
describes the bias between the distributions of HI and the total matter. While the HI bias defined using Eq. \eqref{e_HI_bias} suffers from stochasticity (shot noise), it is closer to observations than results inferred using the alternative definition involving the HI-matter cross-correlation, $b_{\text{HI}}(k) = P_{\text{HI-tot}}(k)/P_{\text{tot}}(k)$ \citep{Sarkar_2016, Castorina_2017}. As shown in Fig. \ref{f_HI_psp_bias}, the large-scale halo bias in CDM at $z=4.94$ attains values $b_{\text{HI}} \approx 2.5$ in redshift space and $b_{\text{HI}} \approx 3.5$ in real space, in agreement with previous work \citep{Sarkar_2016, Villaescusa_2018_2, Wang_2021}. On the smallest (strongly non-linear) scales probed, it increases up to $b_{\text{HI}} \approx 20$. In cFDM cosmologies, the HI bias increases with decreasing axion mass $m$, and for the extreme $m=10^{-22}$ eV model, the bias ratio between cFDM and CDM can reach values above $\approx 2.5$. 

The clustering of the 21 cm signal which is modeled following Sec. \ref{ss_21cm_intro} is shown in Fig. \ref{f_21_psp}. Its magnitude $\propto \Omega_{\text{HI}}^2(z)$ in the post-reionization era is several orders of magnitude lower than before and during reionization, and exhibits a clear trend with decreasing axion mass $m$. The SKA1-Low system noise estimated using \mbox{\scshape{21cmSense}} \normalfont \citep{Pober_2013,Pober_2014} is shown for instrument parameters from Table \ref{t_skalow_params}, assuming an optimistic foreground wedge model in which all $k$ modes inside the primary field of view are excluded \citep{Pober_2014}. Note that we exclude the error arising from sample variance, which predominantly affects larger scales. We find that at $z=4.94$, the 21 cm power spectrum will be detected by this telescope up to scales $k\approx 9$ $h$/Mpc in CDM and the less extreme cFDM models $m\sim 10^{-21}$ eV, while it can be detected up to $k\approx 15$ $h$/Mpc in the more extreme $m=10^{-22}$ eV model.\par

Since we assume that $T_\gamma/T_s \approx 0 $ in our modeling, the relative difference in the 21 cm power spectrum between cFDM and CDM is identical to the relative difference in the HI power spectrum shown in Fig. \ref{f_HI_psp_bias}. We also show the normalized error $\Delta^2_{\text{noise}}/\Delta^2_{21,\text{CDM}}$ for SKA1-Low, representing the $1\sigma$ system noise error on the 21 cm dimensionless power spectrum of the model with CDM. While the more extreme $m=10^{-22}$ eV model can be distinguished from CDM at the $2\sigma$ confidence level across all scales ($k< 80 \ h/$Mpc) resolved by the simulations, SKA1-Low will not be able to discriminate among CDM and the less extreme cFDM models $m\sim 10^{-21}$ eV with $1080$ hours of observations on small scales $k\approx 70$ $h$/Mpc.

\subsection{HI Column Density Distributions}
\label{ss_HI_col}
Another quantity commonly employed to study the distribution of HI in the post-reionization era is the HI CDDF, Eq. \eqref{e_HI_CDDF}. As mentioned in Sec. \ref{ss_estimate_HI}, HI column densities of \Lya \ clouds, LSS, sub-DLAs and DLAs are inferred observationally from quasar spectra.\par
\begin{figure}
\vspace{0.1cm}
\hspace{-0.3cm}
\includegraphics[width=0.5\textwidth]{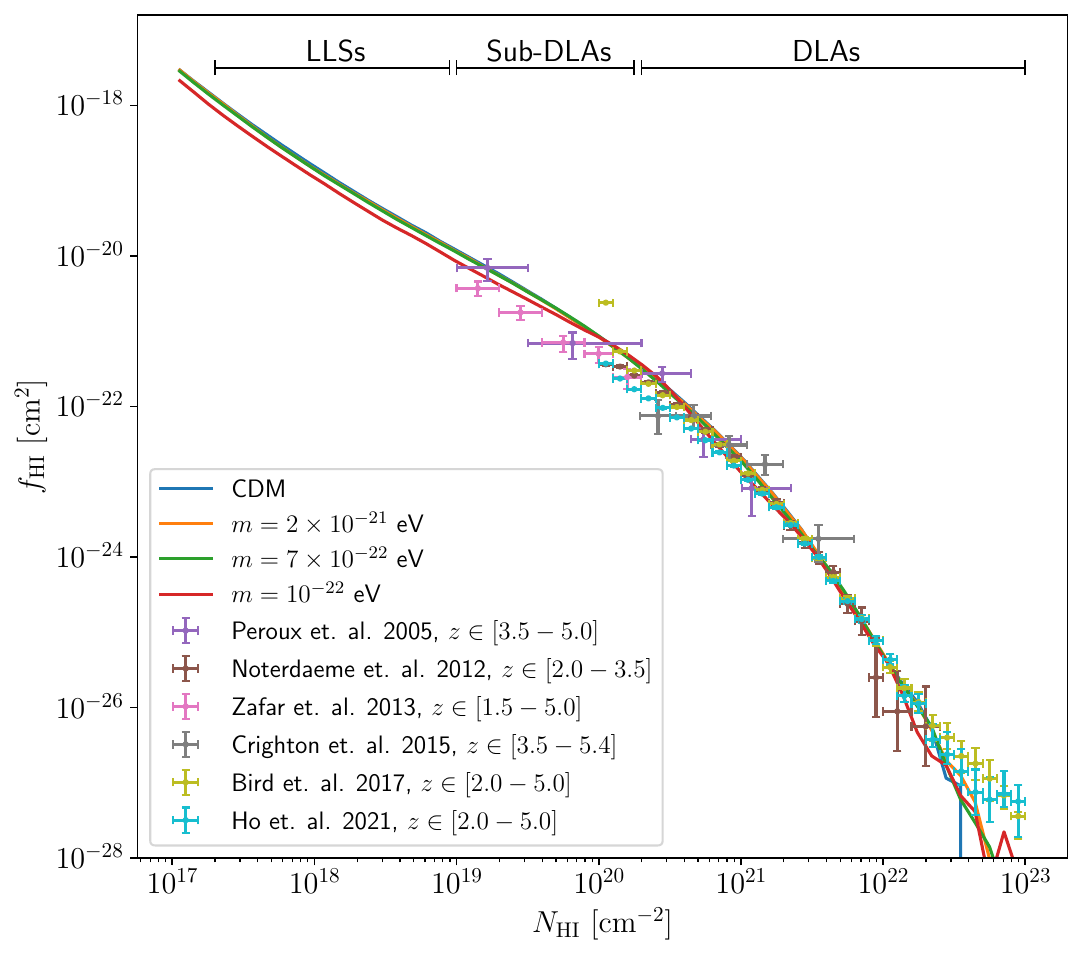}
\caption{Column density distribution function of HI absorbers. Results at $z=3.42$ in CDM and cFDM cosmologies are traced by solid lines. Data from observations are shown with $1\sigma$ error bars: \protect\cite{Peroux_2005} quote \mbox{(sub-)DLA} measurements from ESO UVES; \protect\cite{Noterdaeme_2012} perform a DLA analysis using SDSS DR9 data; \protect\cite{Zafar_2013} quote sub-DLA measurements from ESO UVES; \protect\cite{Crighton_2015} quote results from a Gemini GMOS study of DLAs; \protect\cite{Bird_2017} and \protect\cite{Ho_2021} perform a DLA analysis using SDSS DR12 and DR16Q data, respectively. The redshift ranges in the legend refer to the (sub-)DLA redshifts and not the emission redshifts of the quasars.}
\label{f_NHI}
\end{figure}
To compare mock HI CDDFs to observational datasets, we first assign the HI to gas particles following Sec. \ref{ss_estimate_HI}. Then, the value of the column density $N_{\text{HI}}$ along an arbitrary line of sight (LOS) can be computed following a method from \cite{Villaescusa_2014} which we briefly summarize here. We first project all the gas particle positions onto the XY Cartesian plane, draw LOSs on a regular grid perpendicular to the XY plane from $Z=0$ to $Z=L_{\text{box}}$. For a given LOS, if the minimum distance between any gas particle $i$ and the LOS (i.e. impact parameter $b_i$) is smaller than the smoothing length $h_i$ of the gas particle, then we pick up a contribution based on the integrated density along the path that the LOS intersects the physical size of the gas particle:
\begin{equation}
N_{\text{HI},i} = 2\frac{m_{\text{HI}}}{m_{\text{H}}}\int_0^{l_{\text{max}}}W(r,h_i)\mathrm{d}l,
\end{equation}
where $N_{\text{HI},i}$ is the column density due to particle $i$, having HI mass $m_{\text{HI}}$ and SPH smoothing length $h_i$ while $m_{\text{H}}$ is the mass of the hydrogen atom. The relation between the radius $r$ and the integration variable $l$ is given by $r^2=b^2+l^2$, with $l_{\text{max}}^2=h_i^2-b^2$. Our regular grid consists of $10,000 \times 10,000$ points, i.e. the number of LOSs is $10^8$. In view of our small box size $L_{\text{box}} = 40 \ h^{-1}$Mpc, the probability of encountering more than a single absorber with a large column density $\sim 10^{19}$ cm$^{-2}$ along the LOS is negligible. Our results are thus converged for sub-DLAs and DLAs for which we can, as a result, safely ignore light-cone effects. We repeated the tests mentioned in \cite{Villaescusa_2014} to verify that the grid is fine enough to achieve convergence. For instance, when slicing the box into slabs of different widths and computing the column densities from any of those slabs, the frequency distribution does not change above $\sim 10^{19}$ cm$^{-2}$.\par 
\begin{figure}
\vspace{0.1cm}
\hspace{-0.35cm}
\includegraphics[scale = 0.415]{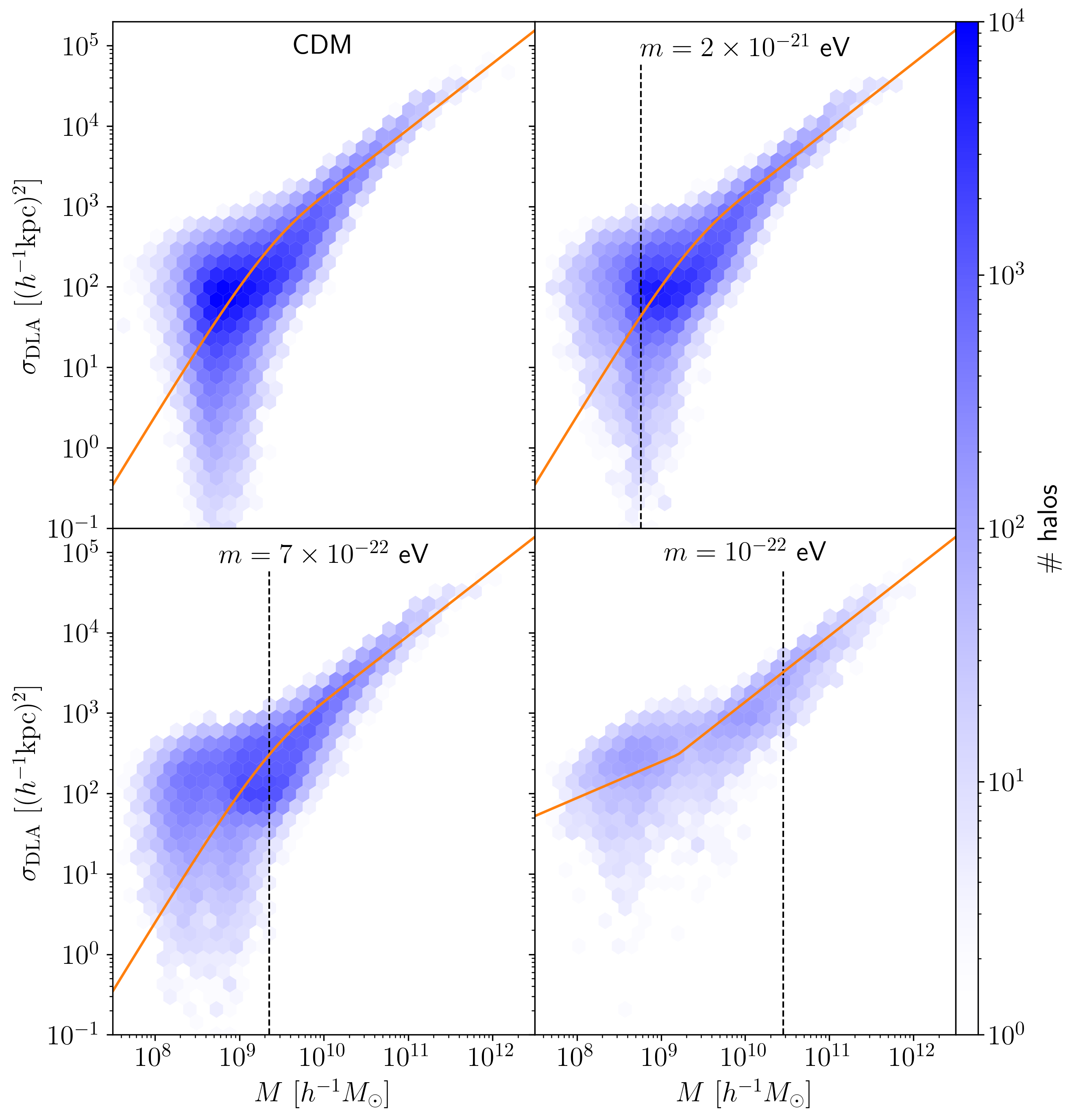}
\caption{DLA cross-section of halos at $z=4.94$. The color of each hexbin is proportional to the number of halos. For the CDM (top left), $m=2\times 10^{-21}$ eV (top right) and $m=7\times 10^{-22}$ eV (bottom left) cosmologies, we fit a power law + exponential function \eqref{e_power_law_cutoff} while for the $m=10^{-22}$ eV (bottom right) cosmology we fit a double power law, Eq. \eqref{e_double_power_law}. The vertical dashed line denotes the half-mode mass $M_{1/2}$ \protect\citep{Marsh_2016}. The small secondary peak in the $m=7\times 10^{-22}$ eV model morphs into a bimodal distribution in the $m=10^{-22}$ eV cosmology.}
\label{f_DLA}
\end{figure}
The inferred frequency distribution of HI column densities in CDM and cFDM cosmologies is shown in Fig. \ref{f_NHI}. We assume a weak redshift dependence of the HI column distribution in the range $z=2-4$, which can be understood in the modeling framework of \cite{Theuns_2021} based on cosmological accretion of gas onto DM halos. Hence, we can overlay results from observations at redshifts $z=[1.5-4.0]$. Since the (sub-)DLAs at $z>4$ are rare and typically inferred from noisy spectra, they mostly contribute to enlarged error bars and we can extend the redshift range to $z=[1.5-5.0]$. We thus add measurements from \cite{Peroux_2005, Noterdaeme_2012, Zafar_2013, Crighton_2015, Bird_2017, Ho_2021} with errorbars to Fig. \ref{f_NHI}. As shown in the most recent measurement paper \citep{Ho_2021}, the trend beyond $z>4$ is in fact still unclear due to statistical uncertainties in the measurements. To better align with the observations, the results in the CDM and cFDM models are shown at $z=3.42$ rather than $z=4.94$ as for the bulk of this paper.\par
 
All simulated cosmologies show good agreement with the observations in the sub-DLA and DLA regimes. The strongest DLA absorbers above $N_{\text{HI}} = 3\times 10^{22}$ cm$^{-2}$ are not modeled well in any of the simulated cosmologies with the predicted CDDF undercutting the measurements, which can be traced to our moderate box size of $L_{\text{box}} = 40 \ h^{-1}$Mpc. The cFDM cosmologies start to diverge from the CDM result in the sub-DLA regime. As a result, with larger data sets and better mean flux measurements, one might be able to put competitive astrophysical constraints on FDM models using sub-DLA surveys. In the LLS and the \Lya \ forest regimes, the abundance of absorbers is expected to differ significantly between CDM and cFDM cosmologies. However, in these regimes we can neither estimate the frequency distribution reliably (as mentioned above) nor is the assumption of a homogeneous UV background valid.\par 
\renewcommand{\arraystretch}{1.6}
\begin{table}
	\centering
	\caption{Best-fit parameters $M_0$ in Eq. \eqref{e_power_law_cutoff} for CDM, $m=2\times 10^{-21}$ eV and $m=7\times 10^{-22}$ eV cosmologies and $M_{\mathrm{break}}$, $\beta$ in Eq. \eqref{e_double_power_law} for the $m=10^{-22}$ eV cosmology across redshifts $z=3.42-4.94$.} 
	\noindent\begin{tabular}{L{1.0cm}L{1.5cm}L{1.5cm}L{1.5cm}}
     & M_0 \ [M_{\odot}/h] & M_{\mathrm{break}} \ [M_{\odot}/h] & \beta \\
	\midrule
	\hline
	z=3.42 & 3.6\times 10^9 & 1.5\times 10^9 & 0.52 \\
	z=3.88 & 2.6\times 10^9 & 1.6\times 10^9 & 0.51 \\
	z=4.38 & 1.9\times 10^9 & 1.7\times 10^9 & 0.48 \\
	z=4.94 & 1.6\times 10^9 & 1.6\times 10^9 & 0.45 \\
	\bottomrule
	\end{tabular}
	\label{t_best_fits}
\end{table}
\renewcommand{\arraystretch}{1}
\subsection{DLA Cross-Sections}
\label{ss_dla_cross_section}
Next, we focus on absorbers with high column densities above $N_{\text{HI,DLA}}=10^{20.3}$ cm$^{-2}$, i.e. DLAs and their cross-sections. The latter are of interest since the observed rate of incidence of DLAs tells us the product of the number density of DLAs times their cross-section \citep{Ribera_2012, Perez_2018}. Both the number density and the cross-section of absorbers is expected to be sensitive to DM physics. In cFDM cosmologies, the high column density of cosmic filaments at $z=2-5$ \citep{Gao_2015} should be reflected in the cross-section distribution, which we now investigate for the first time.\par

We compute the DLA cross-section as follows. For each DM halo, we select the gas particles belonging to the halo and we throw random LOSs within the halo virial radius. After computing the column density for each LOS, the DLA cross-section is readily obtained as
\begin{equation}
\sigma_{\text{DLA}} = \pi R_{\text{vir}}^2\left(\frac{n_{\text{DLA}}}{n_{\text{tot}}}\right),
\end{equation}
where $R_{\text{vir}}$ is the halo virial radius, $n_{\text{DLA}}$ is the number of LOSs with column density above $10^{20.3}$ cm$^{-2}$ and $n_{\text{tot}}$ is the total number of LOSs used. The virial radius is determined as per the \cite{Bryan_1998} spherical collapse result, but we checked that our results do not change if we use a different definition for $R_{\text{vir}}$. We use $40,000$ random LOSs for each DM halo, and find that our results do not change significantly when we increase the number.\par 

\renewcommand{\arraystretch}{1.6}
\begin{table}
	\centering
	\caption{DLA bias $b_{\text{DLA}}$, estimated using Eq. \eqref{e_DLA_bias}, in CDM and cFDM cosmologies across redshifts $z=3.42-4.94$.} 
	\noindent\begin{tabular}{L{1.0cm}L{1.4cm}L{1.4cm}L{1.4cm}L{1.4cm}}
     & \mathrm{CDM} & 2\times 10^{-21} \ \mathrm{eV} & 7\times 10^{-22} \ \mathrm{eV} & 10^{-22} \ \mathrm{eV} \\
	\midrule
	\hline
	z=3.42 & 1.8 & 2.0 & 2.1 & 2.6\\
	z=3.88 & 1.9 & 2.2 & 2.3 & 2.9\\
	z=4.38 & 2.1 & 2.3 & 2.6 & 3.3\\
	z=4.94 & 2.3 & 2.6 & 2.9 & 3.8\\
	\bottomrule
	\end{tabular}
	\label{t_dla_bias}
\end{table}
\renewcommand{\arraystretch}{1}
We show results in Fig. \ref{f_DLA}. We find that the mean DLA cross-section increases with halo mass, which apart from the $m=10^{-22}$ eV cosmology is well fitted by the following function
\begin{equation}
\sigma(M | N_{\text{HI,DLA}},z)=A\left(\frac{M}{h^{-1}M_{\odot}}\right)^{\alpha}\left(1-e^{-(M/M_0)^{\beta}}\right).
\label{e_power_law_cutoff}
\end{equation}
In agreement with the CDM analysis of \cite{Villaescusa_2018_2}, we find that the slope of the cross-section for large halo masses has best-fit value $\alpha = 0.82$ and the characteristic halo mass where the DLA cross-section decreases exponentially has a strong correlation with the overall normalization of the function, $A\cdot M_0 = 14100 \ h^{-3}\text{kpc}^2M_{\odot}$. The cutoff exponent is well approximated by $\beta = 0.85 \cdot \log_{10}(N_{\text{HI,DLA}}/\text{cm}^{-2})-16.35$, the only redshift dependence remaining in $M_0$ and $A$. At $z=4.94$ which is shown in Fig. \ref{f_DLA}, the characteristic halo mass is $M_0 = 1.6\times 10^9 \ h^{-1}M_{\odot}$ in all three cosmologies. In Table \ref{t_best_fits}, we present best-fit values for $M_0$ across redshifts $z=3.42-4.94$, which decreases with redshift implying that less massive halos can host HI at higher redshift \citep[also see][]{Villaescusa_2018_2}.\par 

While a small secondary peak emerges in the cross-section distribution for the $m=7\times 10^{-22}$ eV model, the $m=10^{-22}$ eV cosmology exhibits a bimodal distribution at $z=4.94$. The median cross-section is significantly higher than in the other three cosmologies for halo masses below $M\approx 3\times 10^9 \ M_{\odot}/h$, which is about an order of magnitude below the half-mode mass $M_{1/2}$ \citep{Marsh_2016}. We thus confirm the prediction of \cite{Gao_2015} that the high column density of cosmic filaments in cFDM/WDM cosmologies is reflected in the properties of DLAs. This is also in agreement with cosmic web studies at high redshift, which show that the mean DM overdensity in filaments at $z\approx 5$ is around $30$\% higher in the $m=10^{-22}$ eV cosmology than for CDM \citep{Dome_2023}. The power law + cutoff parametrization of Eq. \eqref{e_power_law_cutoff} thus gives a poor fit for the $m=10^{-22}$ eV model, and instead we adopt a double power law
\begin{equation}
\sigma(M | N_{\text{HI,DLA}},z)=
\begin{cases}
	& A\left(\frac{M}{h^{-1}M_{\odot}}\right)^{\alpha} \ \ \text{if} \ \ M > M_{\mathrm{break}} \\
	& B\left(\frac{M}{h^{-1}M_{\odot}}\right)^{\beta}  \ \ \text{else}.
\end{cases}
\label{e_double_power_law}
\end{equation}
The dependence of $\sigma_{\mathrm{DLA}}$ on cosmology at the high-mass end (i.e. $M > M_{\mathrm{break}}$) is so weak that the normalization $A$ can still be well approximated by $A\cdot M_0 = 14100 \ h^{-3}\text{kpc}^2M_{\odot}$, where $M_0$ is taken from Eq. \eqref{e_power_law_cutoff}. The normalization at the low-mass end $B$ is determined by ensuring that the two power laws intersect at $M_{\mathrm{break}}$. We provide the best-fit values for $M_{\mathrm{break}}$ and $\beta$ in Table \ref{t_best_fits}.\par 

Using the mean DLA cross-section, we can estimate the DLA bias as
\begin{equation}
b_{\text{DLA}}(z|N_{\text{HI,DLA}}) =\frac{\int_0^{\infty}b(M,z)n(M,z)\sigma(M|N_{\text{HI,DLA}},z)\mathrm{d}M}{\int_0^{\infty}n(M,z)\sigma(M|N_{\text{HI,DLA}},z)\mathrm{d}M},
\label{e_DLA_bias}
\end{equation}
where $n(M,z)$ denotes the halo mass function (HMF) and $b(M,z)$ the halo bias. We take the HMF from \cite{Sheth_2002} for CDM and apply the \cite{Schneider_2012} rescaling for cFDM. For the halo bias, we use the fitting formula in \cite{Tinker_2010}, which expresses the bias $b(M,z)$ as a function of the peak height $\nu=\delta_c/\sigma(M)$, where $\delta_c = 1.686$ is the spherical collapse threshold and $\sigma^2(M)$ the linear matter variance. This parametrization provides a reliable fit for the bias of both CDM and cFDM halos when adopting the variance $\sigma^2(M)$ in the respective cosmology \citep{Dunstan_2011}.\par 

In Table \ref{t_dla_bias}, we show the estimated values for the DLA bias in the various cosmologies across redshifts $z=3.42-4.94$, for absorbers with column density $N_{\text{HI}} > N_{\text{HI,DLA}} = 10^{20.3}$ cm$^{-2}$. We find that the DLA bias increases with redshift at the explored post-reionization redshifts. Since the gas in the IGM is denser and the amplitude of the UV background decreases with redshift, this trend is expected for the DLA bias as well as the HI bias. Typically, both biases have similar magnitudes, $b_{\text{DLA}} \simeq b_{\text{HI}}$, following the theoretical arguments in \cite{Castorina_2017} and comparing the values in Table \ref{t_dla_bias} to the $b_{\text{HI}}$ estimates from Fig. \ref{f_HI_psp_bias} at $z=4.94$. In cFDM cosmologies, we find the redshift-independent trend that the DLA bias increases with decreasing axion mass $m$, from $b_{\text{DLA}} = 2.3$ for CDM to $b_{\text{DLA}} = 3.8$ for the $m=10^{-22}$ eV model at $z=4.94$. This trend is primarily a result of the increased halo bias in cFDM cosmologies, as discussed in Sec. \ref{ss_HI_power}.\par

It is useful to compare our estimates of the DLA bias to observations. Through cross-correlating the DLAs with the \Lya \ forest, \cite{Perez_2018} measured a DLA bias factor of $b_{\text{DLA}} = 1.99\pm 0.11$, while using cross-correlations with CMB lensing data \cite{Alonso_2018} and \cite{Lin_2020} measured $b_{\text{DLA}} = 2.6\pm 0.9$ and $b_{\text{DLA}} = 1.37^{+1.30}_{-0.92}$, respectively. All three estimates are for a median DLA redshift of $z_{\mathrm{median}}=2.3$ and agree well with our CDM estimates from Table \ref{t_dla_bias}, if we assume a linear relationship from $z=3.9$ to $z=2.3$ \cite[see e.g.][]{Villaescusa_2018_2}. With more powerful (combined) datasets and a better understanding of the theoretical systematics involved in our use of Eq. \eqref{e_DLA_bias}, the observed DLA bias could be used to put constraints on FDM-like cosmologies.

\section{Mock Radio Maps for SKA-Low}
\label{s_mock_radio}
\renewcommand{\arraystretch}{1.6}
\begin{table}
    \centering
    \begin{tabular}{ll}
         \hline
         Configuration &  \\
         \hline \hline
         Array design & 512 compact core  \\
         \hline
         Station diameter $D$ [m] &  35\\
         \hline 
         Integration time $t_{\mathrm{int}}$ [s] & 60 \\
         \hline
         Bandwidth $B$ [MHz] & 8  \\
         \hline
         Number of channels $N_{\text{chan}}$ & 29/290\textsuperscript{a} \\
         \hline
         Frequency resolution $\Delta \nu$ [kHz] & 272 \\
         \hline
         Number of observing hours/days & 6/180 \\
         \hline
         Redshift/frequency [MHz] & 4.94/238.9 \\
         \hline
         Angular resolution $\theta_{\mathrm{A}}$ & 0.5', 1', 2' \\
         \hline
         Primary beam's FWHM $\theta_{\mathrm{max}}$ & 3.12($\nu/200$ MHz)$^{-1}$ deg \\
         \hline
         Single baseline noise error $\sqrt{\langle |N|^2\rangle}$ [Jy]/[mK] & 0.55/134 \\
         \hline
         \hline
         \multicolumn{2}{l}{\textsuperscript{a}\footnotesize{For imaging: $N_{\text{chan}}=29$; for 21 cm power spectra: $N_{\text{chan}}=290$.}}
    \end{tabular}
    \caption{Summary of SKA1-Low instrument parameters used in this study to obtain the thermal noise sensitivity. The FWHM of the primary beam agrees well with \protect\cite{Kakiichi_2017} and \protect\cite{Hassan_2020}. Natural sensitivities $A_{\mathrm{eff}}/T_{\mathrm{sys}}$ as a function of frequency are tabulated in \protect\cite{Braun_2019}.}
    \label{t_skalow_params}
\end{table}
\renewcommand{\arraystretch}{1}
\renewcommand{\arraystretch}{1.6}
\begin{table*}
	\centering
	\caption{Prospects of detecting the brightest HI peaks using SKA1-Low at $z=4.94$. We provide the peak flux $I_{\mathrm{max}}$, noise rms $\sqrt{\langle |N|^2\rangle}$ (both in $\mu$Jy/beam) and SNR values in CDM and cFDM cosmologies for channel width $\Delta \nu = 272$ kHz, corresponding to a subbox with side length $2.5 \ h^{-1}$Mpc in agreement with the HIGlow modeling of Sec. \ref{ss_higlow}. The peak flux values for the conditional mass of $m=3\times 10^{-22}$ eV (not part of our simulation suite) are obtained by generating $1000$ random samples using HIGlow and searching for the strongest peak among all samples.}
	\noindent\begin{tabular}{L{0.9cm}L{0.9cm}L{0.9cm}L{0.9cm}L{0.9cm}L{0.9cm}L{0.9cm}L{0.9cm}L{0.9cm}L{0.9cm}L{0.9cm}L{0.9cm}L{0.9cm}}
	\toprule
     & \multicolumn{2}{c}{CDM} & \multicolumn{2}{c}{$m=2\times 10^{-21}$ eV} & \multicolumn{2}{c}{$m=7\times 10^{-22}$ eV} & \multicolumn{2}{c}{$m=3\times 10^{-22}$ eV} & \multicolumn{2}{c}{$m=10^{-22}$ eV} & \\
	\cmidrule(r){2-3}\cmidrule(r){4-5}\cmidrule(r){6-7}\cmidrule(r){8-9}\cmidrule(r){10-11}
	\theta_{A} & I_{\mathrm{max}} & \mathrm{SNR} & I_{\mathrm{max}} & \mathrm{SNR} & I_{\mathrm{max}} & \mathrm{SNR} & I_{\mathrm{max}} & \mathrm{SNR} & I_{\mathrm{max}} & \mathrm{SNR} & \sqrt{\langle |N|^2\rangle}\\
	\midrule
	\hline
	0.5' & 0.82 & 0.88 & 0.78 & 0.84 & 0.76 & 0.82 & 0.74 & 0.80 & 0.70 & 0.75 & 0.93\\
	1.0' & 1.39 & 1.81 & 1.31 & 1.70 & 1.27 & 1.65 & 1.23 & 1.60 & 1.17 & 1.52 & 0.77\\
	2.0' & 2.08 & 3.25 & 1.97 & 3.08 & 1.93 & 2.84 & 1.82 & 3.02 & 1.70 & 2.66 & 0.64\\
	\bottomrule
	\end{tabular}
	\label{t_peak_flux}
\end{table*}
\renewcommand{\arraystretch}{1}
Here we investigate the prospects of imaging the HI distribution in the post-reionization era using SKA1-Low. Since SKA maps and summary statistics such as the 21 cm power spectrum are poised to shed light on the nature of DM \citep{Carucci_2015, Bauer_2021}, we study how ML models such as normalizing flows may support such efforts. While in an average region of the Universe, the 21 cm signal may not be strong enough to resolve in imaging, regions with strong HI concentrations are more likely to be resolved. We thus focus on whether SKA1-Low might be able to detect a few individual bright HI peaks and how these peaks link to DM physics.\par

We begin by creating brightness temperature maps from our simulated distribution of HI following Secs. \ref{ss_estimate_HI} and \ref{ss_21cm_intro}. We adopt a channel width of $\Delta \nu = 272$ kHz, corresponding to a subbox with side length $2.5 \ h^{-1}$Mpc in agreement with the HIGlow modeling of Sec. \ref{ss_higlow}. We relate the brightness temperature excess to the specific intensity using the relation
\begin{equation}
I_{\nu}(\hat{\mathbf{r}},\nu) = \frac{2\nu^2}{c^2}k_{\text{B}}\delta T_{\text{b}}(\hat{\mathbf{r}},\nu),
\end{equation}
and project the $I_{\nu}(\hat{\mathbf{r}},\nu)$ field onto a regular grid of resolution $1024\times 1024$ in the XY plane. 

\subsection{Angular Resolution}
The intrinsic resolution of our simulation $\Delta x = L_{\mathrm{box}}/N \approx 40 \ h^{-1}$kpc corresponds to angular scales
\begin{equation}
\Delta \theta = \frac{\Delta x}{D_c(z)} \approx 1.5^{\prime \prime},
\end{equation}
where $D_c(z)$ is the comoving distance to redshift $z=4.94$. On the other hand, the angular resolution of the radio interferometric observation is determined by the maximum baseline $B_{\mathrm{max}}$ via $\theta_A = \lambda/B_{\mathrm{max}}$, where $\lambda$ is the signal wavelength at the observation redshift $z$. Since the maximum baseline of SKA1-Low is $B_{\mathrm{max}} = 65.4$ km corresponding to $\approx 4.0^{\prime \prime}$, SKA1-Low could get close to such angular resolutions. However, due to the compact core layout most of the sensitivity is at the shorter baselines, hence the noise level is lower if one reduces the resolution of the images. In this work, we choose angular resolutions $\theta_A = 0.5^{\prime}, 1^{\prime}, 2^{\prime}$ corresponding to baselines $B_{\mathrm{max}} = 8.6, 4.3, 2.2$ km.\par 

We account for the chosen instrumental resolution by smoothing the $I_{\nu}$ field with a Gaussian window of angular radius $\theta_{A}$, with $\theta_{A}$ being the FWHM of the synthesized beam given in Table \ref{t_skalow_params}. Radio interferometers are not sensitive to the mean value of the specific intensity and can only measure deviations from the mean, hence we recenter the specific intensity field by making the transformation $I_{\nu} \rightarrow I_{\nu} - \langle I_{\nu} \rangle$. Finally, we multiply the $I_{\nu}$ map with the beam solid angle
\begin{equation}
\Delta \Omega = \left(\frac{\pi}{4\ln 2}\right)\theta_{A}^2
\end{equation}
to arrive at the total flux within a single beam, i.e. the resulting image maps will be in units of flux per beam. We repeat the procedure until we find the slice from the simulation box that contains the highest value of $I_{\nu}$, and report results in Table \ref{t_peak_flux}.\par 

\subsection{Thermal Noise}
\label{ss_thermal_noise}
\begin{figure*}
\includegraphics[width=\textwidth]{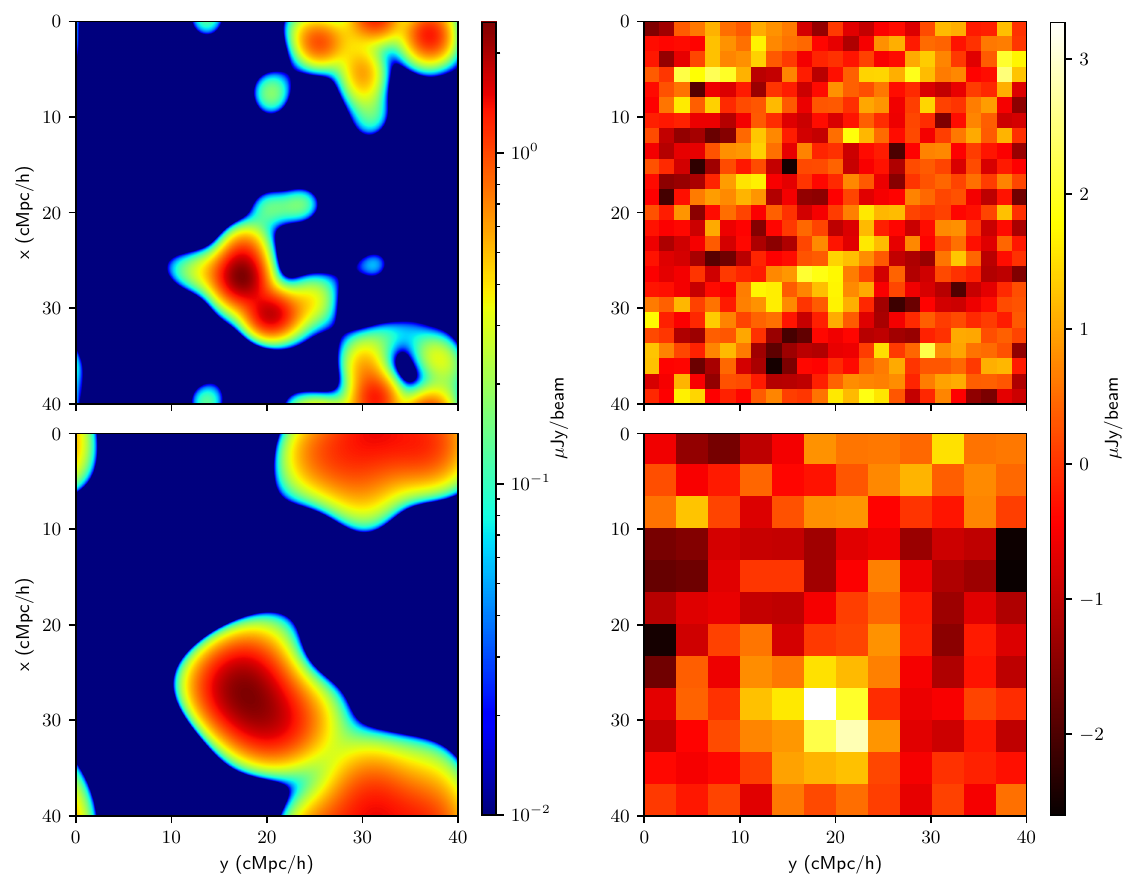}
\caption{Mock radio maps for CDM with a frequency channel of width $\Delta \nu = 272$ kHz and synthesized beam FWHM $\theta_{A} = 1^{\prime}$ (top row) and $\theta_{A} = 2^{\prime}$ (bottom row). Left column: Specific intensity field $I_{\nu}(\hat{\mathbf{r}})$ for $N_{\text{pix}} = 1024$ corresponding to our simulation resolution. Right column: Specific intensity field overlaid with instrumental noise for $N_{\text{pix}} = 12$ and $N_{\text{pix}} = 25$, respectively. We construct the noise map following a four-step procedure outlined in Sec. \ref{ss_thermal_noise}.}
\label{f_mock_radio}
\end{figure*}
To generate noise maps and to estimate the noise levels we largely follow \cite{Hassan_2019}. Every pair of stations in SKA1-Low will record noise along with the visibilities in the $uv$-plane. This noise is uncorrelated between measurements and can be represented by Gaussian random numbers of mean zero and standard deviation \citep[e.g.][]{Ghara_2017}
\begin{equation}
\sqrt{\langle |N|^2\rangle}\mathrm{[Jy]}=\frac{2k_{\mathrm{B}}T_{\mathrm{sys}}}{A_{\mathrm{eff}}\sqrt{\Delta \nu t_{\mathrm{int}}}},
\label{e_noisejy_single_baseline}
\end{equation}
where $t_{\mathrm{int}}$ is the integration time to observe a single visibility at frequency resolution $\Delta \nu$, $A_{\mathrm{eff}}$ is the effective area of each station and $T_{\mathrm{sys}}$ is the system temperature which combines the receiver temperature and the sky temperature. Their combination $A_{\mathrm{eff}}/T_{\mathrm{sys}}$, the natural sensitivity, is frequency dependent and we retrieve the most up-to-date tables from \cite{Braun_2019}. Eq. \eqref{e_noisejy_single_baseline} provides the flux error for a single baseline which we convert to Kelvin using
\begin{equation}
\sqrt{\langle |N|^2\rangle}\mathrm{[K]} = \frac{\lambda^2}{2k_{\mathrm{B}}\Omega}\times \sqrt{\langle |N|^2\rangle}\mathrm{[Jy]},
\end{equation}
where $\Omega = \pi/(4\ln 2)\theta_{\mathrm{max}}^2$ is the full beam area. We present rms noise errors for a single baseline in Table \ref{t_skalow_params}, along with the SKA1-Low instrument parameters.\par 

To generate a 2D thermal noise realization we follow four steps:
\begin{enumerate}
\item We first generate Gaussian random noise (both the real and imaginary parts) with zero mean and rms $\sqrt{\langle |N|^2\rangle}$ from Eq. \eqref{e_noisejy_single_baseline} in the $N_{\mathrm{pix}}\times N_{\mathrm{pix}}$ grid in Fourier space, where $N_{\mathrm{pix}} = \theta_{\mathrm{max}}/\theta_A$.
\item We compute the daily $uv$ coverage $N_{uv}$ which represents the total number of baselines that observe a given $uv$ pixel using \mbox{\scshape{21cmSense}} \normalfont . We account for the presence of multiple baselines in a $uv$ grid point by scaling the noise in the $(i,j)$th pixel by a factor $1/\sqrt{N_{uv}^{ij}}$ at non-zero $uv$ coverage pixels.
\item By averaging over the total number of days observed, we reduce the noise further by a factor of $1/\sqrt{N_{\mathrm{days}}}$.
\item We obtain the real space noise map by inverse Fourier transforming the noise from Fourier space at each frequency channel.
\end{enumerate}
The rms noise values for different values of the angular resolution $\theta_A$ obtained using the above four-step procedure are given in Table \ref{t_peak_flux}. To determine whether there will be enough sensitivity in the maps to identify the cosmological HI using SKA1-Low, we calculate the SNR values of the peak fluxes. While the SNR for $\theta_{A} = 2^{\prime}$ is modest (SNR $> 3$), for $\theta_{A}  = 1^{\prime}$ the SNR is very low (SNR $< 2$). We thus conclude that the imaging prospects of SKA1-Low for $1080$ observing hours and other fiducial parameters listed in Table \ref{t_skalow_params} are moderate at $\theta_{A} = 2^{\prime}$, with a rapidly declining SNR for lower values of $\theta_{A}$. Note that some of the previous works assume a higher number of antenna stations in the SKA1-Low design \citep{Villaescusa_2014, Ghara_2017, Hassan_2019, Hassan_2020}, corresponding to a higher number of baselines and thus lower thermal noise. In particular, for that reason \cite{Villaescusa_2014} arrived at higher peak SNR estimates for SKA1-Low at $z=4$.\par

Accounting for angular resolution and thermal noise, in Fig. \ref{f_mock_radio} we show mock radio maps at $z=4.94$ in CDM for two different angular resolutions, $\theta_{A} = 1^{\prime}$ and $\theta_{A} = 2^{\prime}$. Notably, a prominent peak (SNR$=3.25$) can be observed in the lower half for $\theta_{A} = 2^{\prime}$, corresponding to fluxes per beam around $2 \mu$Jy. However, enhancing the angular resolution comes at a trade-off of increased thermal noise. This effect is depicted in the top panel of Fig. \ref{f_mock_radio}. Here, when adopting $\theta_{A} = 1^{\prime}$, the peak becomes more challenging to distinguish (SNR=$1.81$).

\subsection{HIGlow Interpolation} 
The imaging prospects for cFDM cosmologies are similar to CDM, see peak flux and SNR values in Table \ref{t_peak_flux}. At our fiducial redshift of $z=4.94$, we find a trend of decreasing peak flux $I_{\mathrm{max}}$ for lower values of the axion mass $m$. We could not verify this trend at all redshifts and thus do not believe that it is of physical origin. However, as a proof of concept, we take advantage of this trend and demonstrate the ability of HIGlow to interpolate the latent space of axion masses.\par 

As mentioned in \cite{Hassan_2022}, assuming that HIGlow correctly captures the conditional distribution, it can be used to generate new samples for conditional parameters on which the model has not explicitly been trained. To this end, we generate $1000$ HIGlow samples for a synthetic cosmology with $m=3\times 10^{-22}$ eV (see Sec. \ref{ss_higlow}) and quote the maximum peak flux $I_{\text{max}}$ among all samples in Table \ref{t_peak_flux}. We find that the values are in between the $m=7\times 10^{-22}$ eV and $m=10^{-22}$ eV values inferred from the full-box approach, indicating the success of HIGlow to interpolate its conditional latent space after training.

\section{Conclusions}
\label{s_conclusions}
21 cm intensity experiments with future radio telescopes like the SKA offer a promising source of new astrophysical and cosmological information about the post-reionization era. As such, we can expect competitive constraints on the DM particle mass $m$ and non-cold DM fraction \citep[see e.g.][]{Giri_2022}. In this work, we model the distribution of neutral hydrogen (HI) using high-resolution hydrodynamical $N$-body simulations in CDM and three instances of cFDM cosmologies with axion masses $m=10^{-22}, 7\times 10^{-22}, 2\times 10^{-21}$ eV, adopting the \TNG \ galaxy formation module. Our focus is on the redshift range $z=3.42-4.94$. We demonstrate the ability of the emerging ML technique of normalizing flows to learn the relevant features in the HI distribution and illustrate the detectability of the brightest HI peaks using SKA1-Low. We summarize our main conclusions as follows:

\begin{enumerate}[label={(\alph*)}]
\item We observe a trend of decreasing HI abundance $\Omega_{\text{HI}}$ for smaller axion mass $m$. Comparing to a range of recent measurements, we show that both CDM and the less extreme cFDM models $m\sim 10^{-21}$ eV are in good agreement therewith. However, the statistical and systematic errors on the measurements are too large to put competitive constraints on FDM using $\Omega_{\text{HI}}$ directly.
\item \label{p_HI_bias} The HI bias $b_{\text{HI}}^2(k) = P_{\text{HI}}(k)/P_{\text{tot}}(k)$ as inferred from the ratio of the HI and total matter power spectra increases with decreasing axion mass, as previously establish. We identify that it is the higher halo bias in cFDM cosmologies \citep[e.g.][]{Dunstan_2011} that leads to increased HI clustering. This is also reflected in the amplitude of the 21 cm power spectrum which is up to $300$\% higher for the $m=10^{-22}$ eV model than for CDM at $z=4.94$, similar to \citep{Carucci_2015}. We find that SKA1-Low will be able to discriminate among the $m=10^{-22}$ eV model and CDM at the $2\sigma$ confidence level on all scales $k<80 \ h/$Mpc resolved by the simulations.
\item The weak dependence of the HI frequency distribution $f_{\text{HI}}(N_{\text{HI}}, X)$ on redshift in the post-reionization era allows to compare to a range of DLA and sub-DLA measurements covering $z=[1.5-5.0]$. These show good agreement with predictions from all our investigated CDM and cFDM models. We show that in the sub-DLA regime, cFDM cosmologies start to exhibit a lower abundance of HI absorbers compared to CDM.
\item DLA cross-sections exhibit a simple power law + low-mass exponential cutoff trend for cFDM models $m\sim 10^{-21}$ eV that is well known for CDM models \citep{Villaescusa_2018_2}. For more extreme models with $m\sim 10^{-22}$ eV, we show that a double power law parametrization provides a better fit. The high median cross-section at the low-mass end can be traced to the high column density of cosmic filaments \citep{Gao_2015}. Using the mean DLA cross-section, we estimate the DLA bias $b_{\text{DLA}}$ and recover the well-known result of increasing DLA bias with increasing redshift. In addition, we find the redshift-independent trend that the DLA bias \textit{increases} with decreasing axion mass $m$, from $b_{\text{DLA}} = 2.3$ for CDM to $b_{\text{DLA}} = 3.8$ for the $m=10^{-22}$ eV model at $z=4.94$. This trend is in agreement with Point \ref{p_HI_bias} and is again a result of the increased halo bias in cFDM cosmologies.
\item The prospects of imaging the brightest HI peaks with SKA1-Low at the fiducial redshift of $z=4.94$ are moderate (SNR $>3$) for angular resolutions $\theta_A \approx 2$, with a rapidly declining SNR for lower values of $\theta_{A}$. Our peak SNR estimates are lower than for the comparable study of \citep{Villaescusa_2014} who assumed $911$ stations for SKA1-Low as opposed to $512$.
\item We demonstrate the ability of normalizing flows, a generative ML model introduced by \cite{Agnelli_2010}, to capture intricate structures resulting from non-linear physics. We adapt the HIGlow framework \citep{Hassan_2022} to facilitate the \textit{conditional} generation of HI maps with varying axion masses. We use two validation metrics, HI mass PDFs and HI power spectra. As proof of concept, we showcase the ability of HIGlow to interpolate its latent space of axion masses to \textit{predict} the peak flux value for a synthetic cosmology with $m=3\times 10^{-22}$ eV.
\end{enumerate}
As we advance into the systematics-dominated era and witness the construction of PB-scale radio telescopes such as SKA, in the quest for DM it is imperative to improve the modeling of HI distributions in various CDM and non-CDM cosmologies. In the post-reionization era that we focus on, it is easier to disentangle DM signatures from astrophysical processes, which are poorly understood during the epoch of reionization and cosmic dawn. Observations during \citep[e.g.][]{Castellano_2023} and after reionization \citep[e.g.][]{Irsic_2023} are thus complementary and will increase the robustness of DM constraints. However, a more comprehensive analysis of astrophysical processes is needed even in the post reionization era to swiftly explore a wide array of galactic and stellar feedback effects that can be marginalized over. Advances in cosmological simulations realizing thousands of astrophysical models \citep{Navarro_2021} and seminumerical modeling techniques \citep{Giri_2022} signify a major stride in the pursuit of this goal.

\section{Acknowledgements}
It is a pleasure to thank Oscar O'Hara for enriching conversations on the intricacies of radio interferometers, particularly in the context of the future Square Kilometer Array. We express our sincere gratitude to Roy Friedman for his support in bolstering the numerical stability of the HIGlow model. We also appreciate insightful discussions with Emily Wickens. TD acknowledges support from the Isaac Newton Studentship and the Science and Technology Facilities Council (STFC) under grant number ST/V50659X/1. AF is supported by the Royal Society University Research Fellowship. The simulations were performed under DiRAC project number ACSP253 using the Cambridge Service for Data Driven Discovery (CSD3), part of which is operated by the University of Cambridge Research Computing on behalf of the \href{https://dirac.ac.uk}{STFC DiRAC HPC Facility}. The DiRAC component of CSD3 was funded by BEIS capital funding via STFC capital grants ST/P002307/1 and ST/R002452/1 and STFC operations grant ST/R00689X/1. DiRAC is part of the National e-Infrastructure.

\section{Data Availability}
\label{s_data_availability}
The \TNG \ galaxy formation module specifications are publicly accessible at \href{https://www.tng-project.org/}{https://www.tng-project.org/}. Post-processing scripts are made available upon reasonable request.

\bibliographystyle{mnras}
\bibliography{refs}

\appendix

\section{HIGlow Implementation Details}
\label{a_cond_nflows}
Normalizing flows are a class of generative models with rich applications in the field of probability density estimation \citep{Dinh_2015} and can thus be used to learn HI distributions. While unconditional normalizing flows are in general easier to train, the conditional normalizing flow framework HIGlow \citep{Friedman_2022} with the below implementation details proves sufficiently stable to learn HI distributions across the latent space of axion masses.\par 

Our training dataset consists of pairs $\{(x_i,c_i)\}_{i=1}^N$, where each $x_i$ represents a set of normalized HI maps with dimensions $64\times 64$, and the $c_i$ correspond to the logged axion mass parameters $\log_{10}(m_a)$ that serve as conditional parameters for the model. By virtue of the Schr\"odinger-Vlasov correspondence \citep{Mocz_2018}, the dynamics of FDM and N.B. cFDM converge to CDM dynamics when the de Broglie wavelength $\lambda_{\text{dB}}$ is significantly smaller than the spatial resolution $\Delta x$ of our HI maps. We impose
\begin{equation}
\frac{\lambda_{\text{dB}}}{\Delta x} = \frac{h}{m_{\text{CDM}} \varv} \frac{1}{\Delta x} < 10^{-3}
\end{equation}
and find $m_{\text{CDM}} = 10^{-19}$ eV, i.e $-19$ as the CDM conditional parameter, assuming a local characteristic velocity of DM (in the Madelung formulation) of $\varv \sim 20$ km/s \citep{Dome_2022}. Our full training set thus contains $40,000$ samples. Our objective now is to approximate the true underlying probability density function (PDF) $p(x|c)$ using a function $p_{\boldsymbol{\theta}}(x|c)$.\par

\subsection{Loss Function}
A normalizing flow
\begin{equation}
f = f_1 \circ f_2 \circ ... \circ f_k
\end{equation}
is a composition of invertible functions $(f_i)_{i=1}^k$ that maps a simple known distribution (such as a spherical multivariate Gaussian noise $y \sim \mathcal{N}(\mathbf{0},\mathbf{1})$) to a more complicated distribution (such as the probability distribution of our HI maps). Note that the functions all implicitly depend on the parameters $\boldsymbol{\theta}$ of the model. We define intermediate variables
\begin{equation}
h_i = f_1 \circ f_2 \circ ... \circ f_i(x,c)
\end{equation}
with $h_0 = (x,c)$ and $h_k = y$. The key to turning this into an efficient computational model is to choose the functions $f_i$ to be simple with known analytical inverses and Jacobian determinants. This allows us to write the loss function of our model as the Kullback-Leibler (KL) divergence between the underlying PDF $p$ and the model's estimated PDF $p_{\boldsymbol{\theta}}$:
\begin{equation}
L(\boldsymbol{\theta}) = D_{KL}(p||p_{\boldsymbol{\theta}}),
\end{equation}
which we minimize with respect to $\boldsymbol{\theta}$. The KL divergence between two distributions $p$ and $q$, 
\begin{equation}
D_{KL}(p || q) = \int_{-\infty}^{\infty}p(x)\log \left( \frac{p(x)}{q(x)} \right) dx,
\end{equation}
measures a statistical distance between distributions. For a large number of samples from the true distribution, we get the following estimate for the loss function \citep{Papamakarios_2021}:
\begin{equation}L(\boldsymbol{\theta})=-\frac{1}{N}\sum_{i=1}^N \log p_{\boldsymbol{\theta}}(x_i|c_i).
\end{equation}
Hence our loss function is just the negative log-likelihood. Applying the standard change of variables formula for PDFs, we get
\begin{align}
    \log p_{\boldsymbol{\theta}}(x|c) &=  \log p_{\boldsymbol{\theta}}(y) + \log \left|\det \left(\frac{dy} {d(x,c)}\right)\right| \notag \\
    &= \log p_{\boldsymbol{\theta}}(y) + \sum_{i=1}^k \log \left|\det \left(\frac{dh_i} {dh_{i-1}}\right)\right|. 
    \label{e_NLL1}
\end{align}
In Eq. \eqref{e_NLL1}, the first term is the standard log-likelihood of a multivariate Gaussian. Since we choose simple functions to make up the flow, the Jacobians and their determinants in the second term have simple analytical expressions. Now another great advantage of normalizing flows becomes evident: the exact likelihoods are known and worked with throughout the training. By contrast, other generative models like generative adversarial networks (GANs) and variational autoencoders (VAEs) do not allow us to extract the exact likelihoods. By inverting the flow, normalizing flows thus provide a tractable high-dimensional likelihood for constraint forecasting and parameter inference in astrophysics and cosmology \citep[e.g.][]{Zamudio_2019, Bevins_2022, Hassan_2022}.\par
\renewcommand{\arraystretch}{1.6}
\begin{table}
    \centering
    \begin{tabular}{ll}
        \hline
          Layer & Forward Operation  \\
         \hline \hline
          Conditional affine injector& $x \mapsto g_s(c) \circ x + g_t(c), x=(x_a, x_b)$  \\
         \hline
         Invertible convolution& $x \mapsto \mathbf{W}x, x=(x_a, x_b)$\\
         \hline
         Conditional affine coupling& $x_a \mapsto x_a, x_b \mapsto f_s(x_a,c) \circ x_b + f_t(x_a, c)$\\
         \hline
    \end{tabular}
    \caption{Three layers making up a conditional flow block. $g_s, g_t$ are simple neural networks, and $f_s, f_t$ are some simple well behaved functions. In particular, a sigmoid function is used for $f_s$ because a requirement of the affine coupling block is that this function should not take a value of zero \protect\citep{Behrmann_2021}. Note how the model is made conditional by introducing the conditional parameter $c$ into the affine injector and affine coupling layers.}
    \label{t_flowblocktable}
\end{table}
\renewcommand{\arraystretch}{1}
Our model is naturally a generative one. Once trained, we can start with a Gaussian noise value $y$, the axion mass parameter $c$ and generate a sample HI map $x$ by applying the inverse flow $g$:
\begin{equation}
x = g(y,c) = f^{-1}(y,c).
\label{e_invert}
\end{equation}

\subsection{Architecture}
The challenge of designing a normalizing flow architecture is choosing the functions $f_i$ that make up the flow. There are various different existing architectures such as RealNVP \citep{Dinh_2017} and NICE \citep{Dinh_2015}. However, we use a Glow architecture \citep{Kingma_2018} which has been modified to be conditional \citep{Friedman_2022}. The building block of the unconditional Glow model is a flow step, which is the composition of an actnorm layer, an invertible $1\times 1$ convolution, and an affine coupling layer. Multiple flow steps are composed to make a flow block, and multiple flow blocks make up the whole model. The flow blocks are connected using a multi-scale architecture \citep{Dinh_2017} which splits the data in between flow blocks. This approach allows the use of the spatial structure of the HI maps during training. To make Glow conditional, the actnorm and affine coupling layers are modified in a simple way to incorporate conditioning on parameters. Table \ref{t_flowblocktable} shows the three key layers that make up a conditional flow block.\par 
\renewcommand{\arraystretch}{1.6}
\begin{table}
    \centering
    \begin{tabular}{ll}
         \hline
         Hyperparameter & Value \\
         \hline \hline
         Flow blocks & 6  \\
         \hline
         Flow steps per block &  12\\
         \hline 
         Hidden width & 20 \\
         \hline
         Learning rate & $3 \times 10^{-4}$  \\
         \hline
         Sigmoid rescaling $s$ & $0.5$ \\
         \hline
         Epochs & 1000 \\
         \hline
    \end{tabular}
    \caption{Hyperparameter values used for our model. Both the conditional affine coupling as well as the affine injector have one hidden layer of width given above. The sigmoid rescaling parameter $s$ refers to the constant in \mbox{Eq. \eqref{e_rescaled_sig}}.}
    \label{t_hparams}
\end{table}
\renewcommand{\arraystretch}{1}
The inverse flow and thus sample generation can be prone to numerical instabilities \citep{Behrmann_2021}, even if the loss curve plateaus as expected and training is successful. To stabilize the inverses against small numerical errors, we choose small Lipschitz constants $L_i$ for the functions that constitute the flow and their respective inverses $g_i$. By rearranging the definition of Lipschitz continuity,
\begin{equation}
\frac{\norm{g_i(x_1)-g_i(x_2)}}{\norm{x_1 - x_2}} \leq L_i,
\end{equation}
a smaller Lipschitz constant $L_i$ thus implies that $g_i$ has small gradients (if differentiable). An upper bound on the Lipschitz constant for the inverse of an affine coupling block (see Table \ref{t_flowblocktable}) is established in \cite{Behrmann_2021}. The bound depends linearly on the reciprocal of the sigmoid function and its derivative. However, the reciprocal of the sigmoid function grows exponentially to $\pm$infinity for large inputs close to $0,1$, respectively. Thus, the Lipschitz constant becomes potentially unbounded, and so do the numerical errors associated with this layer. The sigmoid function
\begin{equation}
S(x) = \frac{1}{1 + \exp(-x)},
\end{equation}
appears as the function $f_s$ in the affine coupling layer of Glow (see Table \ref{t_flowblocktable}). We thus modify the affine coupling layer to use a linearly rescaled sigmoid function, 
\begin{equation}
\Tilde{S}(x) = (1-s)S(x) + s,
\label{e_rescaled_sig}
\end{equation}
where $s$ is a constant we set. $\Tilde{S}$ takes values on the interval $(s,1)$ rather than $(0,1)$ and therefore assures that the relevant Lipschitz constant is bounded from above for large enough values of $s > 0$.\par

Another approach to stabilize the inverse flow consists of image clipping, i.e. constraining pixel values to fall within predefined maximum and minimum ranges after each flow block in the generative reverse flow. However, this method leads to the loss of crucial image features with each successive clipping operation. Consequently, we opt for the use of the rescaled sigmoid given in Eq. \eqref{e_rescaled_sig} to maintain fidelity without compromising essential image characteristics.\par

The key hyperparameters used for training are shown in Table \ref{t_hparams}. To allow the maximum amount of splitting \citep{Dinh_2017} of the $64^2$ images throughout the flow, we employ $6$ flow blocks. This enables effective utilization of spatial structures at various scales during training. $12$ flows per block and a hidden layer width of $20$ make sure the model has enough parameters to learn the key features without over- or underfitting. On average, training took $6$ hours per $100$ epochs on an NVIDIA A100 GPU. The final model that is presented in this work was trained for $1000$ epochs.\par

\subsection{Data Pre-Processing}
We pick $10,000$ random $64^3$ subcubes from the full 3D HI distribution in each of the four cosmologies. The subcubes are then projected to get $64^2$ HI maps. These $64^2$ maps have a comoving length of $2.5\ h^{-1}$Mpc; this length scale is chosen to balance the requirement of a large number of samples while keeping the individual maps sufficiently large to capture DM imprints. Using projected maps leads to more stable and less computationally expensive training.\par

If $X$ denotes the unnormalized map from our training data, we first take the logarithm of each map, $x = \log_{10}(X)$. We also sigmoid-normalize our data onto the interval $(-1,1)$ by
\begin{equation}
\Tilde{x} = \frac{2}{1+\exp(-\hat{x})} - 1,
\label{e_norm}
\end{equation}
where $\Bar{x}$ and $\sigma_x$ is the mean and standard deviation of the whole training dataset, respectively, and $\hat{x} = (x-\Bar{x})/\sigma_x$.

\label{lastpage}
\end{document}